\documentclass{pasj00}
\usepackage{color}

\begin{document}
\SetRunningHead{Arimoto et al.}{Spectral Lag Relations}
\Received{2009/12/27}
\Accepted{2010/02/17}

\title{Spectral Lag Relations in GRB Pulses Detected with HETE-2}

\author{
M. \textsc{Arimoto}\altaffilmark{1},
N.  \textsc{Kawai}\altaffilmark{1},
K.  \textsc{Asano}\altaffilmark{1},
K. \textsc{Hurley}\altaffilmark{2}, 
M. \textsc{Suzuki}\altaffilmark{3}, 
}

\author{
Y. E. \textsc{Nakagawa}\altaffilmark{4},
T. \textsc{Shimokawabe} \altaffilmark{1},
N. V. \textsc{Pazmino} \altaffilmark{1},
R. \textsc{Sato}\altaffilmark{5},
M. \textsc{Matsuoka}\altaffilmark{3},
}
\author{
A. \textsc{Yoshida}\altaffilmark{6},
T. \textsc{Tamagawa}\altaffilmark{4}, 
Y. \textsc{Shirasaki}\altaffilmark{7},
S. \textsc{Sugita}\altaffilmark{6},
I. \textsc{Takahashi}\altaffilmark{6},
}

\author{
J.-L. \textsc{Atteia}\altaffilmark{8},
A. \textsc{Pelangeon}\altaffilmark{8},
R. \textsc{Vanderspek}\altaffilmark{9},
C. \textsc{Graziani}\altaffilmark{11},
G. \textsc{Prigozhin}\altaffilmark{9},
}
\author{
J. \textsc{Villasenor}\altaffilmark{9},
J. G. \textsc{Jernigan}\altaffilmark{2},
G. B. \textsc{Crew}\altaffilmark{9},
T. \textsc{Sakamoto}\altaffilmark{12},
G. R.  \textsc{Ricker}\altaffilmark{9},
}

\author{
S. E. \textsc{Woosley}\altaffilmark{16},
N. \textsc{Butler}\altaffilmark{2},
A. \textsc{Levine}\altaffilmark{9},
J. P. \textsc{Doty}\altaffilmark{9,10},
T. Q. \textsc{Donaghy}\altaffilmark{11},
}
\author{
D. Q. \textsc{Lamb}\altaffilmark{11},
E. \textsc{Fenimore}\altaffilmark{15},
M. \textsc{Galassi}\altaffilmark{15},
M. \textsc{Boer}\altaffilmark{17},
J.-P. \textsc{Dezalay} \altaffilmark{13},
}

\author{
J.-F. \textsc{Olive} \altaffilmark{13},
J. \textsc{Braga} \altaffilmark{18},
R. \textsc{Manchanda} \altaffilmark{19},
and
G. \textsc{Pizzichini} \altaffilmark{14}
}

\altaffiltext{1}{Department of Physics, Tokyo Institute of Technology,
2-12-1 Ookayama, Meguro-ku, Tokyo 152-8551}
\email{arimoto@hp.phys.titech.ac.jp}
\altaffiltext{2}{Space Sciences Laboratory, University of California,
Berkeley, California, 94720-7450}
\altaffiltext{3}{JAXA, 2-1-1 Sengen, Tsukuba, Ibaraki, 305-8505}
\altaffiltext{4}{RIKEN, 2-1 Hirosawa, Wako Saitama 351-0198}
\altaffiltext{5}{Institute of Space and Astronautical Science/JAXA, Sagamihara, Kanagawa, 229-8510, Japan}
\altaffiltext{6}{Department of Physics and Mathematics, Aoyama Gakuin
University, \\5-10-1 Fuchinobe, Sagamihara, Kanagawa 229-8558}
\altaffiltext{7}{National Astronomical Observatory of Japan, Osawa,
Mitaka, Tokyo, 181-8588}
\altaffiltext{8}{LATT, Universit{\'e} de Toulouse, CNRS, 14 Avenue E. Belin, 
31400 Toulouse, France}
\altaffiltext{9}{Center for Space Research, MIT, 77 Vassar Street,
Cambridge, Massachusetts, 02139-4307, USA}
\altaffiltext{10}{Noqsi Aerospace, Ltd., 2822 South Nova Road, Pine,
Colorado, 80470, USA}
\altaffiltext{11}{Department of Astronomy and Astrophysics, University of
Chicago, \\5640 South Ellis Avenue, Chicago, Illinois 60637, USA}
\altaffiltext{12}{Goddard Space Flight Center, NASA, Greenbelt,
Maryland, 20771, USA}
\altaffiltext{13}{Centre d'Etude Spatiale des Rayonnements, 
9 Avenue du Colonel Roche, Toulouse, 31028, France}
\altaffiltext{14}{INAF/IASF Bologna, Via Gobetti 101, 40129 Bologna,
Italy}
\altaffiltext{15}{Los Alamos National Laboratory, P.O.Box 1663, Los
Alamos, NM, 87545, USA}
\altaffiltext{16}{Department of Astronomy and Astrophysics, University
of California at Santa Cruz, \\477 Clark Kerr Hall, Santa Cruz, California,
95064, USA}
\altaffiltext{17}{Observatoire de Haute-Provence (CNRS/OAMP), Saint Michel
l'Observatoire, France}
\altaffiltext{18}{Instituto Nacional de Pesquisas Espaciais, Avenida Dos
Astronautas 1758, \\ Sa{\~o} Jos{\'e} dos Campos 12227-010, Brazil}
\altaffiltext{19}{Department of Astronomy and Astrophysics, \\Tata
Institute of Fundamental Research,  Homi Bhabha Road, Mumbai, 400-005,
India } 


%


\maketitle

\begin{abstract}
Using a pulse-fit method,
we investigate the spectral lags
between the traditional gamma-ray band (50$-$400 keV) and the X-ray band
(6$-$25 keV) for 8 GRBs with known redshifts
(GRB 010921, GRB 020124, GRB 020127, GRB 021211, GRB
 030528, GRB 040924, GRB 041006, GRB 050408) detected with
the WXM and FREGATE instruments aboard the {\it HETE-2} satellite.
We find several relations for the individual
GRB pulses between the spectral lag and other observables, such as the luminosity,
pulse duration, and peak energy $E_{\rm peak}$.
The obtained results are consistent with those for BATSE, indicating that the
BATSE correlations are still valid at lower energies (6$-$25 keV).
Furthermore, we find that the photon energy dependence for the spectral lags
can reconcile the simple curvature effect model.
We discuss the implication
of these results from various points of view.
\end{abstract}

\section{Introduction}
Gamma-Ray Bursts (GRBs) are the most energetic explosions in the universe.
Past studies  have found that GRBs consist of
ultra-relativistic outflows with collimated jets
at cosmological distances. 
However, it is not clear how the central engine forms and how the
 electrons or protons are accelerated in shocks and photons are
radiated.
In addition,  GRBs are quite important as candidates for distance-indicators.
Owing to their very intense brightness, GRBs can be a powerful tool to
 measure distances in the high redshift universe.

One of the characteristics of GRB prompt emission is the spectral lag,
which is the time delay in the arrival of lower-energy emission
 relative to higher-energy emission.
The previous analyses have been done using a sample of many
BATSE GRBs between typical energy bands 
 25$-$50 keV and 100$-$300 keV, using both CCF (cross correlation function;
e.g., \cite{2000ApJ...534..248N})
 and peak-to-peak difference (e.g., \cite{2008ApJ...677L..81H}).
An anti-correlation between the spectral lag and
the luminosity exists for the BATSE GRBs above 50 keV energies.
Since we can obtain the intrinsic luminosity of GRBs from the
 lag-luminosity relation once we measure the spectral lag, 
the distance of the GRBs can be derived from the
 observed flux. But it is not clear whether the relation is valid in wider 
energy bands.
In addition, from the results of \cite{2008ApJ...677L..81H}, it is shown that 
the spectral lag characterizes
each {\it pulse} rather than the entire burst.

From the theoretical point of view
(e.g., \cite{2004ApJ...617..439Q}), the rise phase timescale may be
 responsible for the intrinsic pulse width,
while the decay phase timescale may be determined by 
geometrical effects (e.g., the curvature effect).
The curvature effect (\cite{2002A&A...396..705Q}, \cite{2005MNRAS.362.1085Q},
\cite{2006MNRAS.367..275L}) arises from relativistic effects in a sphere 
expanding with a high bulk Lorentz factor $\Gamma$ = 1/(1 - $\beta^2$)$^{1/2}$
 $\sim$ 100. Because of the
curvature of the emitting shell, 
there will be a time delay between the photons emitted
simultaneously in the comoving frame from different points on the surface.
However, \cite{2007PASJ...59..857Z} showed that
the curvature effect alone is not enough to explain
energy-dependent pulse properties obtained from
the systematic analysis of
lag and temporal evolution.
Alternative models are the off-axis model proposed by 
\cite{2001ApJ...554L.163I}　and 
the time-evolution of shock propagation (\cite{1998MNRAS.296..275D},
\cite{2003MNRAS.342..587D}, \cite{2009A&A...498..677B}) may also
reproduce the spectral lag and the lag-luminosity relation.
Thus, it is not clear that either the curvature effect or
other effects cause the spectral lag.
While the curvature effect should necessarily affect the pulse profile,
the time-evolution of shock propagation or off-axis model
strongly depends on unknown model parameters.

 In this paper, in order to unveil the properties of the spectral lag
for each pulse, we investigate the {\it HETE-2} sample with a wider
energy  range especially at the low-energy end ($>$2 keV) than the BATSE sample.
In sections 2 and 3
we explain the sample selection
and the pulse-fit method.
In section 4, we describe the result of the obtained relations between
the spectral lag and other observables, and we 
discuss a detailed energy dependence for
the spectral lag in section 5. Finally, we briefly comment on
the future prospects in section 6.

\section{{\it HETE-2} Sample and Selection}
{\it HETE-2} had two scientific instruments on-board which are relevant to our
study:
the FREnch GAmma-ray TElescope (FREGATE), which gave the trigger for GRBs and 
performed spectroscopy over a wide energy range 
(6$-$400 keV); and the Wide-field
X-ray Monitor (WXM), which was the key instrument to localize GRBs to
$\sim$10$^\prime$, and sensitive to the 2$-$25 keV energy range, lower than
the FREGATE one. The instruments have two types of data.  The survey 
data were recorded with fixed energy bands and time resolution whenever the
instruments were on.
The time-tagged
data were produced with a fixed duration (several minutes) when 
the instruments were triggered by bursts.
From the time-tagged data, we can produce light curves in arbitrary energy 
bands, while  the BATSE detector in general created light curves only in
fixed energy bands (Although the BATSE detector actually has time-tagged 
data, many BATSE GRBs are not fully covered due to the limitation of the  memory
size for the time-tagged data.).

We perform the spectral-lag analysis using a sample of 8 GRBs detected by {\it HETE-2} with known or estimated redshifts for the study of the lag-luminosity relation in section \ref{sec:lag-relation}.
 Our selection criteria for the GRB samples are the following: 1) 
$T_{90}$ $>$ 2 s,
where $T_{90}$ is the observed duration including 90\% of the total
observed counts,
and 2) time-tagged data are available.
For the latter, we note that the time-tagged data were lost for some 
bursts due to downlink
problems or invalidation of the instruments (e.g., GRB030328, GRB030329 etc.).
For these bursts, since
the available energy band is too coarse for the survey data (e.g., 6$-$40 keV, 6$-$80 keV, and 32$-$400
keV for FREGATE), we cannot conduct a detailed study of  the spectral lag.
 For the analysis in section \ref{sec:lag-relation}, we use the FREGATE
 instrument alone 
because off-axis photon events were partially coded 
and the number of events detected by the WXM instrument was often small,
while the FREGATE instrument  
detected more photons compared to those of the WXM instrument due to its
relatively large effective area ($\sim$150 cm$^2$); not all the selected GRBs
 have enough photons to perform the analysis in the WXM energy band.

In addition, for studying the detailed energy dependence of the 
spectral lag for individual GRB pulses 
 in section \ref{sec:detailed_ene_dep}, 
we add 2 GRBs without known redshifts having sufficiently 
non-overlapped pulses to the sample.  In this analysis, we use not only the 
FREGATE instrument but also the WXM instrument, because some GRBs have
good enough statistics detected by the WXM instrument.
Here, since there are not good statistics in the
multiple energy bands for GRB 020124 and GRB 041006, 
 we exclude the GRBs from the sample.

\begin{figure}[h]
\centering
\includegraphics[width=9cm]{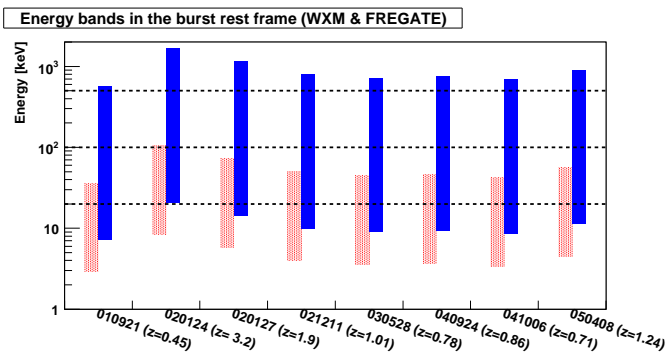}
\caption{Energy
 bands in the burst rest frame for the selected GRBs. 
The red dotted bars represent the WXM bands
 and the blue solid ones represent the FREGATE bands. The adopted energy ranges
 are the range between the dashed lines (20$-$100 keV and 100$-$500 keV).
}
\label{fig:adoptedEnergyband}
\end{figure}

We show the list of 10 GRBs in Table 
\ref{table:HETEsample}  and the energy bands in the burst rest frame 
which are covered by the WXM and FREGATE instruments for the selected GRBs with
known redshifts in Fig. \ref{fig:adoptedEnergyband}.

\begin{table}[h]
  \caption{GRB samples}\label{tab:HETEsamples}
  \begin{center}
    \begin{tabular}{|c|c|c|}
     \hline
     GRB & redshift & Reference \\ \hline
     010921 & 0.45 & \cite{2001GCNDjorgovski} \\ \hline	
     020124 & 3.20 & \cite{2003ApJ...597..699H} \\ \hline
     020127 & 1.9$^1$  & \cite{2007ApJ...660..504B} \\ \hline	
     021211 & 1.01 & \cite{2003GCN..1785....1V} \\ \hline	
     030528 & 0.78 & \cite{2005AA...444..425R} \\ \hline	
     030725 & -  & \cite{2005AA...439..527P} \\ \hline
     040924 & 0.86 & \cite{2004GCN..2800....1W} \\ \hline	
     041006 & 0.72 & \cite{2005ApJ...626L...5S} \\ \hline	
     050408 & 1.24 & \cite{2005GCN..3201....1B} \\ \hline	
     060121 & - & \cite{2006ApJ...648L..83D} \\ \hline	
    \end{tabular}
   \label{table:HETEsample}
  \end{center}
   $^1$: this is a possible value estimated from the afterglow investigation
 and spectral energy distribution.
\end{table}

\section{Method}
Each GRB pulse is fitted with a four-parameter pulse model
\citep{2005ApJ...627..324N},  if $t$ $>$ $t_{\rm start}$
\begin{eqnarray}
 I(t) &=&  A \lambda \exp{ (-\tau_{1}/(t-t_{\rm start}) - 
 (t-t_{\rm start})/\tau_{2} )} + B(t)  \\ 
 &=&  A \lambda \exp  (-\tau_{1}/(t + \tau_{\rm peak} - t_{\rm peak}) \nonumber \\
&&- (t + \tau_{\rm peak} - t_{\rm peak})/\tau_{2} )  + B(t) 
\label{eq:Norris2005_2}
\end{eqnarray}
 and if $t$ $<$ $t_{\rm start}$, $I(t) = B(t)$ ,
where $I$ is the intensity, $t$ is the time after the trigger, $\tau_1$ and $\tau_2$ are  the pulse rise and pulse decay constants, $\lambda \equiv \exp{\left(2 (\tau_1/\tau_2)^{1/2}\right) }$, $t_{\rm peak}$ is the time of the pulse's maximum intensity $A$,
$t_{\rm start}$ is the start time, $\tau_{\rm peak} \equiv (\tau_1\tau_2)^{1/2}$ is the peak
time from the start time $t_{\rm start}$, so that
$t_{\rm peak} = t_{\rm start} + \tau_{\rm peak}$,  and $B(t)$ is the background 
function (we utilize a constant or linear function). 
In Eq. \ref{eq:Norris2005_2}, $t_{\rm peak}$ is treated as a primary
fitting parameter in order to estimate the uncertainty in $t_{\rm peak}$
directly in the fitting procedure.

The time $t_{\rm start}$ is the formal onset time and in some cases 
$t_{\rm start}$ is not indicative of the visually apparent onset
time. Especially in the case of $\tau_1	 \gg 1 $ s, $t_{\rm start}$
is extremely far from the peak of the pulse.
Here, as described in \cite{2005ApJ...627..324N}, we introduce an effective
onset time $t_{\rm eff}$ arbitrarily defined as the time when the pulse reaches 0.01
times the peak intensity.
Furthermore, the values of  $t_{\rm eff}$  are different in different energy
bands. For {\it HETE-2} GRBs, the statistics of GRBs are not as good as
those of BATSE because, e.g., the effective area of the FREGATE
detector ($\sim$ 150 cm$^{2}$) is lower by a factor of $\sim$ 10 than that of
the BATSE detector ($\sim$ 2000 cm$^2$).
This causes $t_{\rm eff}$ to be scattered in different energy bands due
to the uncertainties in the determination of $\tau_1$ and $\tau_2$.
To avoid this, we adopt an onset time of the ``bolometric'' light-curve
profile, $t_{\rm eff}^\prime$, derived by fitting the light curve in the 6
$-$ 400 keV band, which corresponds to the entire FREGATE-energy band. 
 The adoption of $t_{\rm eff}^\prime$ is
supported by \cite{2009ApJ...705..372H}. They
showed that the onset of GRB pulses occurs simultaneously
across all energy bands.
Thus, we define $T_{\rm peak}$ = $t_{\rm peak} - t_{\rm eff}^\prime$ 
 in this paper.
The corresponding uncertainties are calculated using the error propagation formula.

Spectral peak lags are defined as the difference between the
maximum-intensity times
in different energy bands as
\begin{eqnarray}
\tau_{\rm lag} \equiv t_{\rm peak, low} - t_{\rm peak, high},
\end{eqnarray}
 where ``low'' and ``high'' represent the low and high energy bands,
respectively.
Another measurable pulse property is
the pulse duration
$w \equiv 3\tau_{2} { (1 +4\sqrt{\tau_1/\tau_2}/3) }^{1/2}$ defined as
the time intervals where intensities are equal to $e^{-3} I(t_{\rm peak})$.

\section{Relation between the Spectral Lag and Other Parameters}\label{sec:lag-relation}

In this paper, we adopt two sets of energy bands in order to calculate the
spectral lag between the two divided bands.
The first one is 6$-$25 keV and 50$-$400 keV in the observer's frame.
Although in the previous study by BATSE the energy bands between 25$-$50 keV and
100$-$300 keV had been adopted, we adopt the lower energy band ($<$25
keV) and test if the same relation (e.g., lag-luminosity relation) is 
established or not.
Furthermore, for all the previous studies of the spectral lag the energy bands
refer to the {\it observer's} frame. However if the spectral lag is a
characteristic property of GRBs, it is better to derive the
spectral lags between the energy bands in the {\it burst rest} frame.
 The {\it HETE-2} time-tagged data have an advantage for such an analysis, 
compared with the BATSE detector.
Thus we adopt energy bands 20$-$100 keV and 100$-$500 keV in the
burst rest frame to be covered by the FREGATE instrument.
The adopted energy bands are shown as horizontal lines in Fig. \ref{fig:adoptedEnergyband}.

\begin{figure}
  \begin{center}
    \FigureFile(80mm,80mm){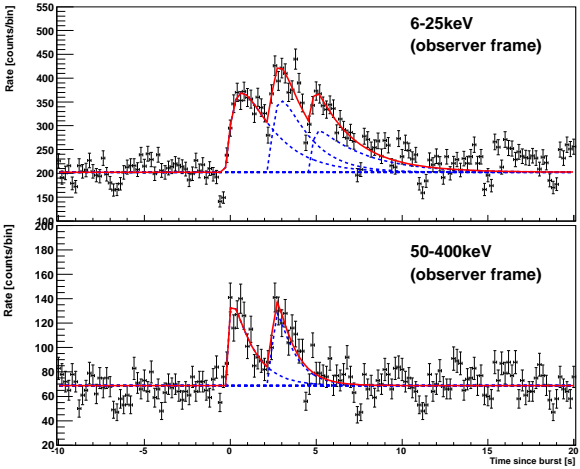}
  \end{center}
  \caption{Pulse fit for GRB 050408 in the 6-25 keV and 50-400 keV bands in the observer's frame.}\label{fig:pulsefit050408obs}
\end{figure}

Here we use pulses
that satisfy the following requirements: the significance of the spectral lag
$\sigma_{\rm sig} > 1.5 $ and positive lag $\tau_{\rm lag} > 0$,
where
\begin{eqnarray}
\sigma_{\rm sig} &\equiv& \tau_{\rm lag} / \sigma_{\rm lag},
\label{eq:definition_spectrallag} \\
\sigma_{\rm lag} &\equiv& \left(\sigma_{\rm peak, low}^2 +\sigma_{\rm peak, high}^2\right)^{1/2}.
\end{eqnarray}
The value $\sigma_{\rm lag}$ represents the 1-$\sigma$ uncertainty of 
$\tau_{\rm lag}$ and $\sigma_{\rm peak}$ is defined as the 1-$\sigma$ uncertainty of
$t_{\rm peak}$. 
For the negative lag, its significance is low ($\sigma_{\rm lag} < 1$) and the 
number of pulses having a negative lag is very small, and we do not
take it into account.
One of the pulse-fitted results is shown in Fig. \ref{fig:pulsefit050408obs} 
(GRB 050408) using the $\chi^2$ fitting routine. For the fitting,
we fit pulses to  make the obtained values of $\chi^2$/d.o.f. (degree of freedom)  
 reasonable ($\sim$1).

\subsection{Observer's frame}

\begin{figure*}[htbp]
 \begin{center}
  \begin{minipage}[xbt]{7.5cm}
   \resizebox{7.5cm}{!}
   {\includegraphics[angle=0]{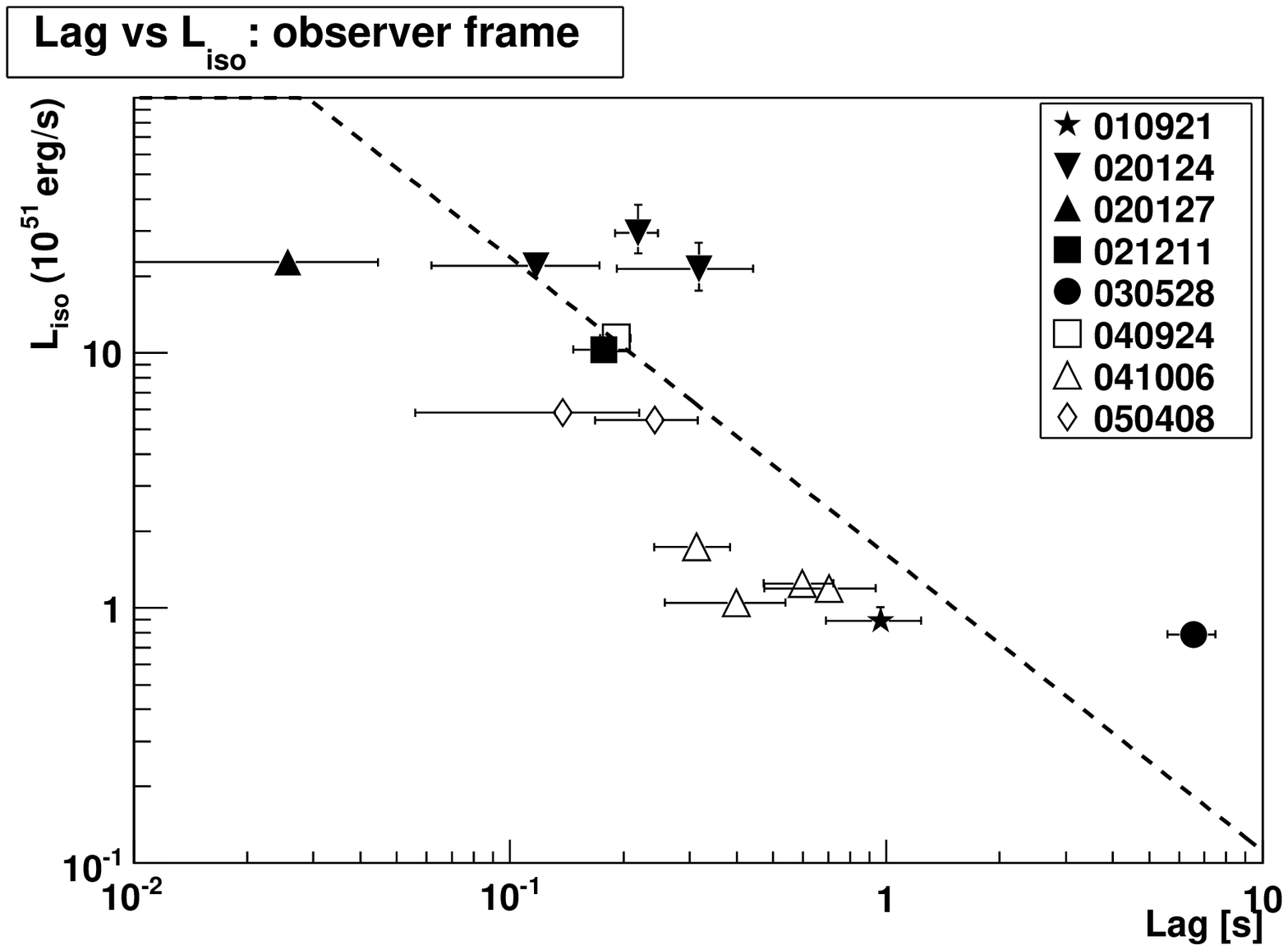}}
  \end{minipage}
  \begin{minipage}[xbt]{7.5cm}
   \resizebox{7.5cm}{!}
   {\includegraphics[angle=0]{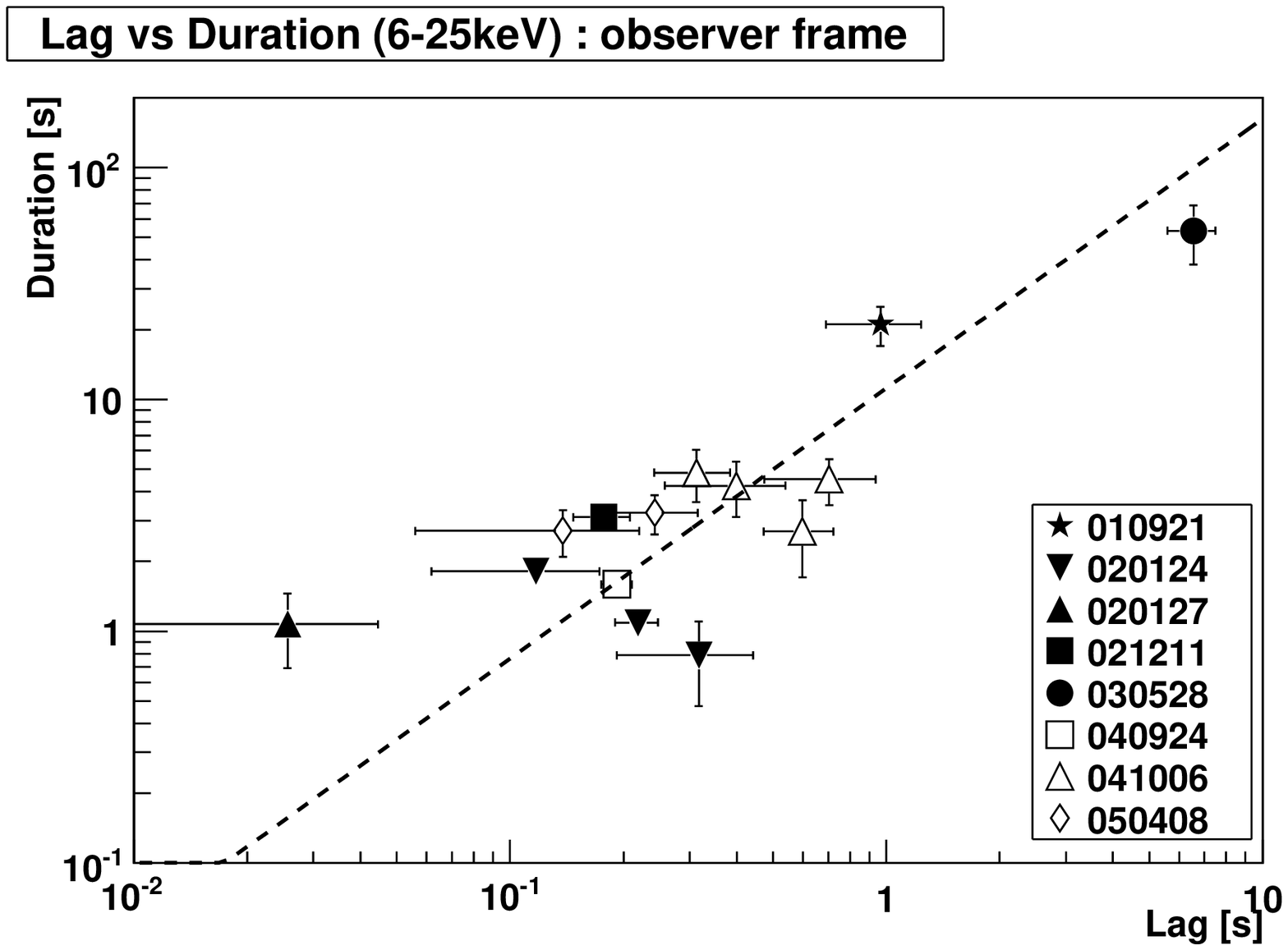}}
  \end{minipage}
  \begin{minipage}[xbt]{7.5cm}
   \resizebox{7.5cm}{!}
   {\includegraphics[angle=0]{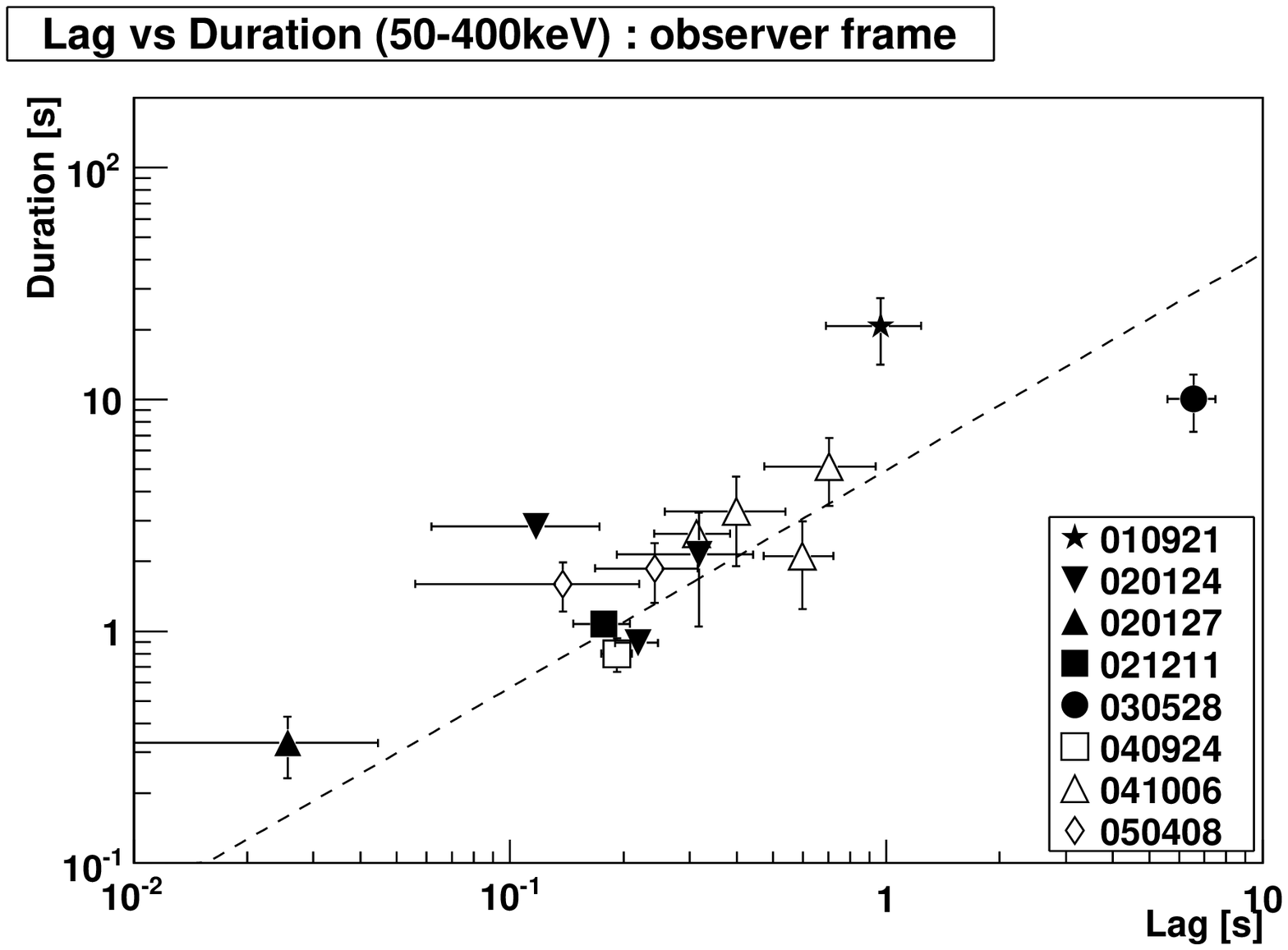}}
  \end{minipage}
  \begin{minipage}[xbt]{7.5cm}
   \resizebox{7.5cm}{!}
   {\includegraphics[angle=0]{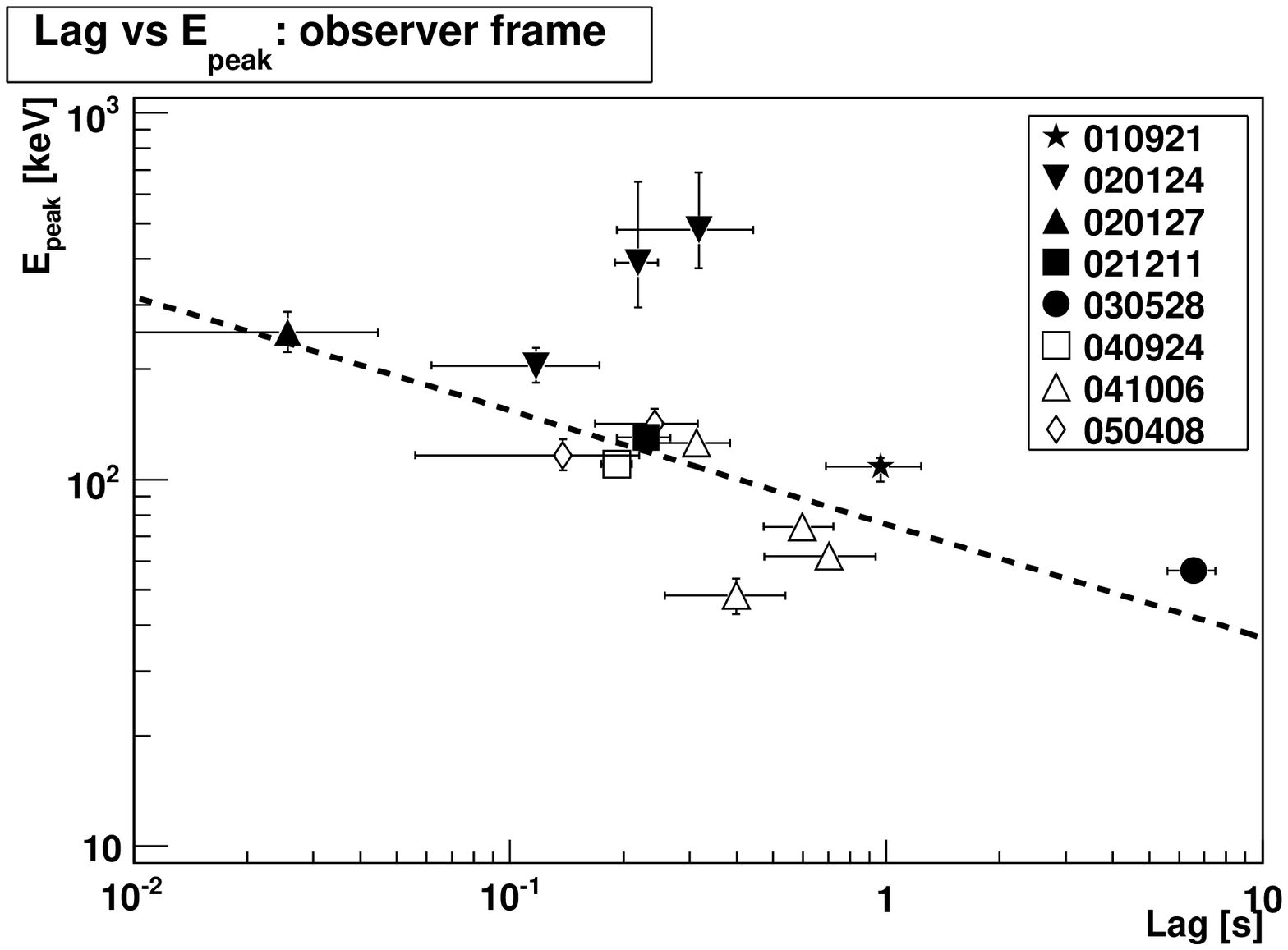}}
  \end{minipage}
 \end{center}
 \caption{Scatter plot (observer's frame) of
 spectral lag vs. luminosity (top left), duration for the low-energy band 
(top right), duration for the high-energy band (bottom left), and $E_{\rm peak}$
(bottom right). 
The dashed lines represent the best-fit functions.
 }
 \label{fig:correlation_lagvsluminosity}
\end{figure*}

First, we show the scatter plots of the spectral lag $\tau_{\rm lag}$ and 
luminosity $L_{\rm iso}$ in the top left panel of Fig. 
\ref{fig:correlation_lagvsluminosity} where the spectral lag is calculated
 in the observer's frame.
 The luminosity in this paper
is defined as the average luminosity over the pulse FWHM timescale.
As shown in this figure, we can see an anti-correlation
 between the spectral lag and the luminosity.
Here, for the correlation coefficient, we adopt the 
{\it Spearman rank-order} correlation test.
Furthermore, to estimate the correlation coefficient on the basis of the
spectral lag's confidence level, we perform a Monte Carlo simulation.
Since we have already obtained $\tau_{\rm lag}$,  $\sigma_{\rm lag}$,
 the luminosity and its uncertainty, we can generate a {\it 
pseudo } plot based on the specific probability distributions, that is,
make a plot  similar to 
the top left panel of Fig. \ref{fig:correlation_lagvsluminosity} with random
number seeds.
Then we calculate the value of the correlation coefficient $R$ for
 the generated {\it pseudo } plot using the {\it Spearman
rank-order} correlation test.
Finally we repeat the same procedure 10000 times with different random number seeds.
As we obtain the histogram of the correlation coefficient, we regard the 
1-$\sigma$ width as the 1-$\sigma$ confidence level.
We adopt this method in the following analysis.

For the lag-luminosity relation in the observer's frame,
we obtain the correlation coefficients as $R$ = -0.79$^{+0.16}_{-0.05}$ 
with a chance probability of 7.7 $\times$
10$^{-4}$ at the most probable value.
The best-fit functional form is $\log(L_{51}) = A_1 + B_1 \log(\tau_{\rm lag}) $
with $A_1 = -0.79 \pm 0.04, B_1 = -1.16 \pm 0.07$;  the reduced chi-square 
is 133.2/12
in the observer's frame. 
 Although there are large scatters in the data from the 
best-fit line,  the lag-luminosity relation holds even for the low energy
band ($<$ 25 keV).
 While our results reconfirm the lag-luminosity relation
previously reported,
our spectral lag index (-1.2) is slightly smaller than
that of \cite{2008ApJ...677L..81H} (index $\sim$-0.6).
The slight difference seems to come from the following:
(1) the different timescales to estimate the luminosity
(BATSE used 256 ms, while we adopt the pulse FWHM timescale),
(2) the small numbers of GRB samples for both
\cite{2008ApJ...677L..81H} and  {\it HETE-2},
and (3) the difference in the adopted energy band and/or the instrumental 
response between \cite{2008ApJ...677L..81H} and ours.
Thus, the slight difference in the power-law index
of the correlations between  \cite{2008ApJ...677L..81H} and our results 
is not surprising.

We show the scatter plots of the spectral lags and the durations 
at low and high energies in the top right and bottom 
left panels of Fig. 
\ref{fig:correlation_lagvsluminosity}, respectively. 
We find correlations between the spectral lags and durations in 
both the 6$-$25 keV and 50$-$400 keV bands.
The best-fit
functional form of these relations is $\log(w_{\rm low}) = A_2 + B_2 \log(\tau_{\rm lag}) $
with $A_2 = 1.05 \pm 0.06, B_2 = 1.16 \pm 0.09$ and the reduced chi-square is 
43.0/12  in the 6$-$25 keV band and 
$\log(w_{\rm high}) = A_3 + B_3 \log(\tau_{\rm lag}) $
with $A_3 = 0.69 \pm 0.05, B_3 = 0.94 \pm 0.08$, and 
45.1/12  in the 50$-$400 keV band.
The correlation 
coefficients are $R$ = 0.66$^{+0.10}_{-0.14}$ and $R$ = 0.74$^{+0.06}_{-0.14}$,
and the corresponding chance probabilities at the most probable values are
1.0 $\times$ 10$^{-2}$ and 2.5 $\times$ 10$^{-3}$
 in the 6$-$25 keV and 
50$-$400 keV bands, respectively.
The low
chance probabilities assure the tight correlations.
These results are almost consistent with those of
\cite{2008ApJ...677L..81H} (index 0.85).

Finally, the scatter plot of the spectral lag and  peak-time $E_{\rm peak}$
 is
shown in the bottom right panel of Fig. \ref{fig:correlation_lagvsluminosity} in 
the observer's frame .
The best-fit functional form is $\log(E_{\rm peak}) = A_4 + B_4 \log(\tau_{\rm lag}) $
with $A_4 = 1.88 \pm 0.02, B_4 = -0.31 \pm 0.02$ and the reduced
chi-square is 97.8/12.
The correlation coefficient is $R$ = -0.66$^{+0.15}_{-0.08}$ 
 with chance probability 1.0 $\times$
10$^{-2}$.
Thus, we obtain a  possible anti-correlation between $E_{\rm peak}$ and the spectral lag,
though the dependence of the 
spectral lag on $E_{\rm peak}$ is relatively weak (index $\sim$ -0.3 ) 
compared with the other parameters.

\subsection{Burst Rest frame}
\begin{figure*}[htbp]
 \begin{center}
  \begin{minipage}[xbt]{75.0mm}
   \resizebox{75.0mm}{!}
   {\includegraphics[angle=0]{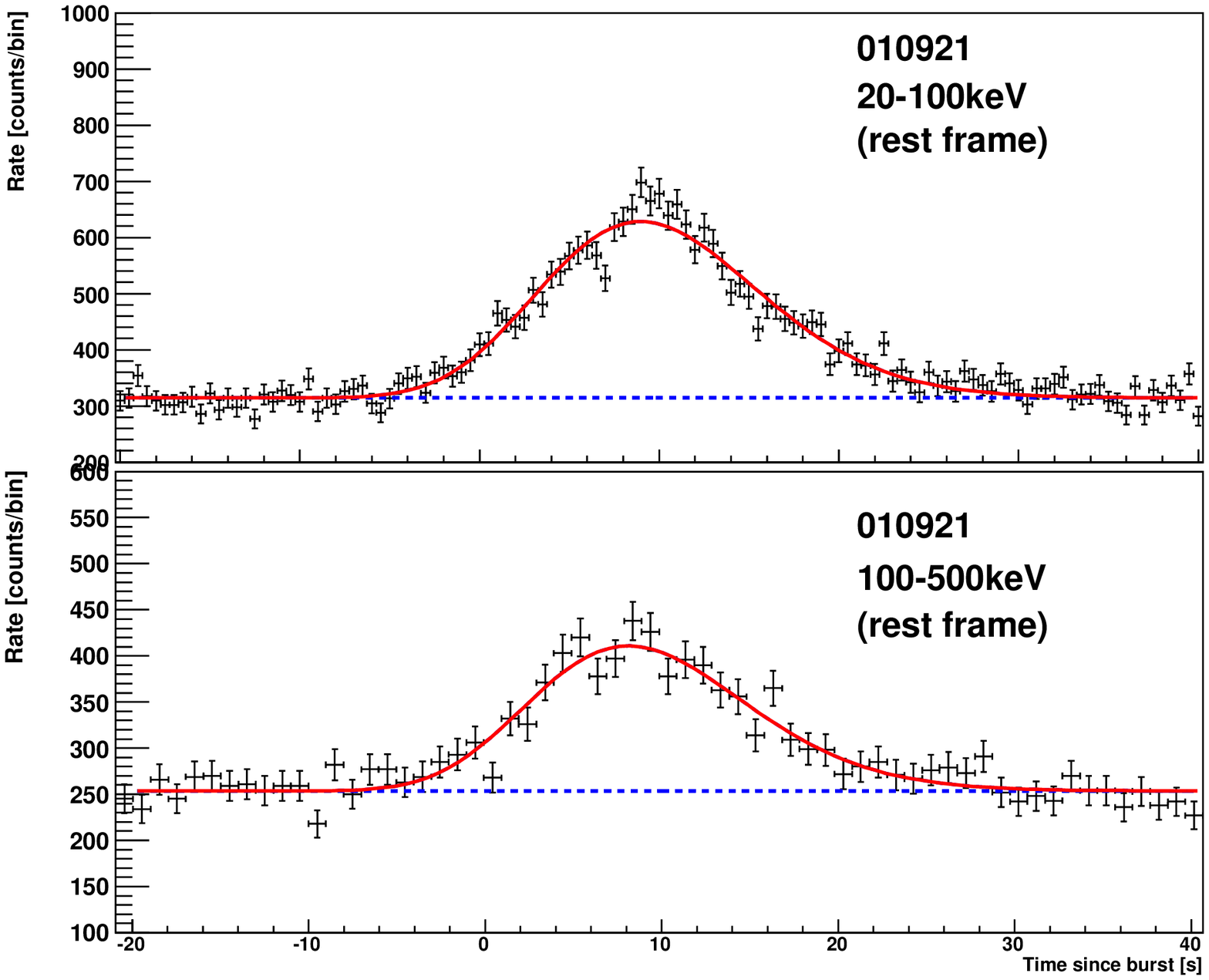}}
  \end{minipage}
  \begin{minipage}[xbt]{75.0mm}
   \resizebox{75.0mm}{!}
   {\includegraphics[angle=0]{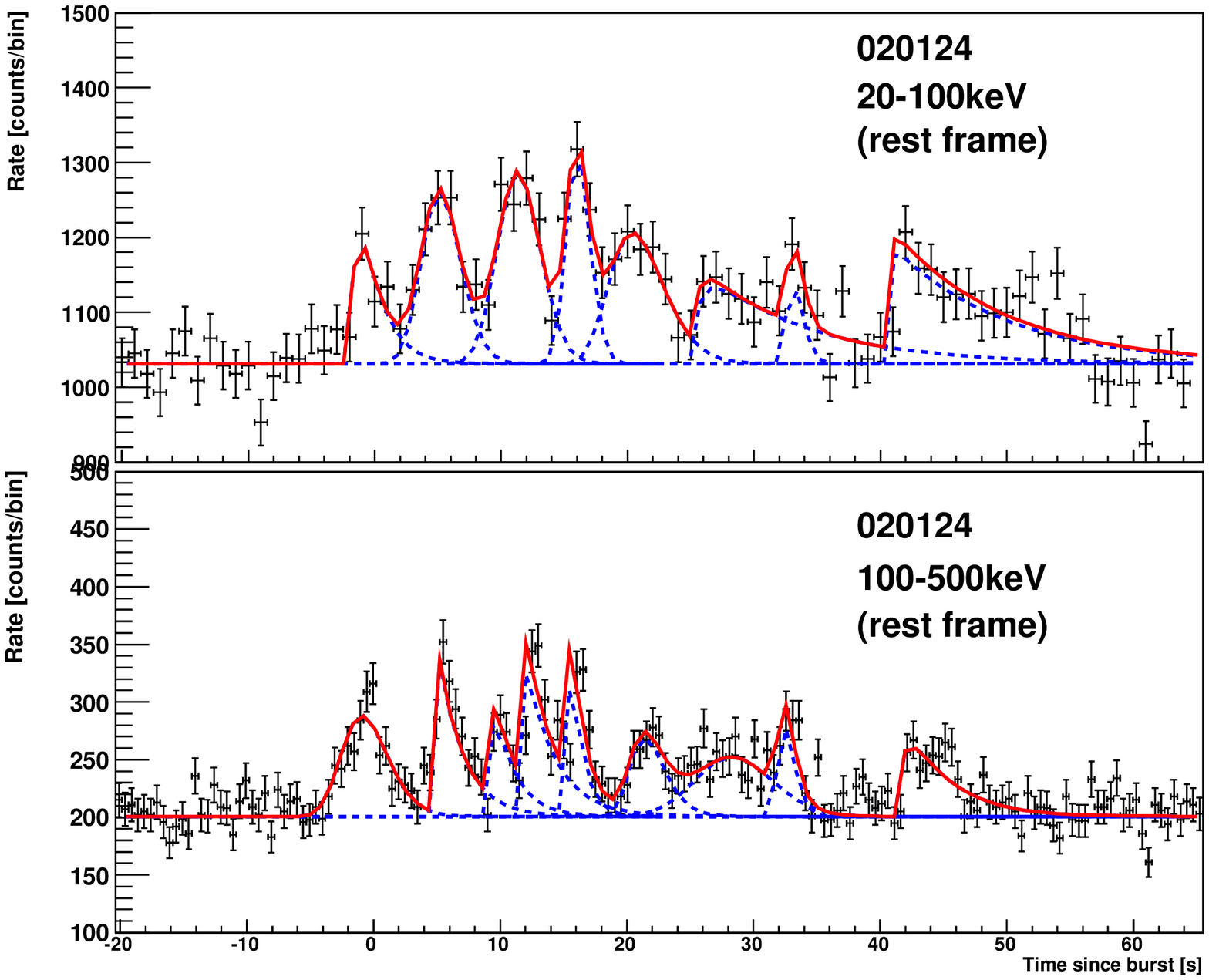}}
  \end{minipage}
 \end{center}
 \begin{center}
  \begin{minipage}[xbt]{75.0mm}
   \resizebox{75.0mm}{!}
   {\includegraphics[angle=0]{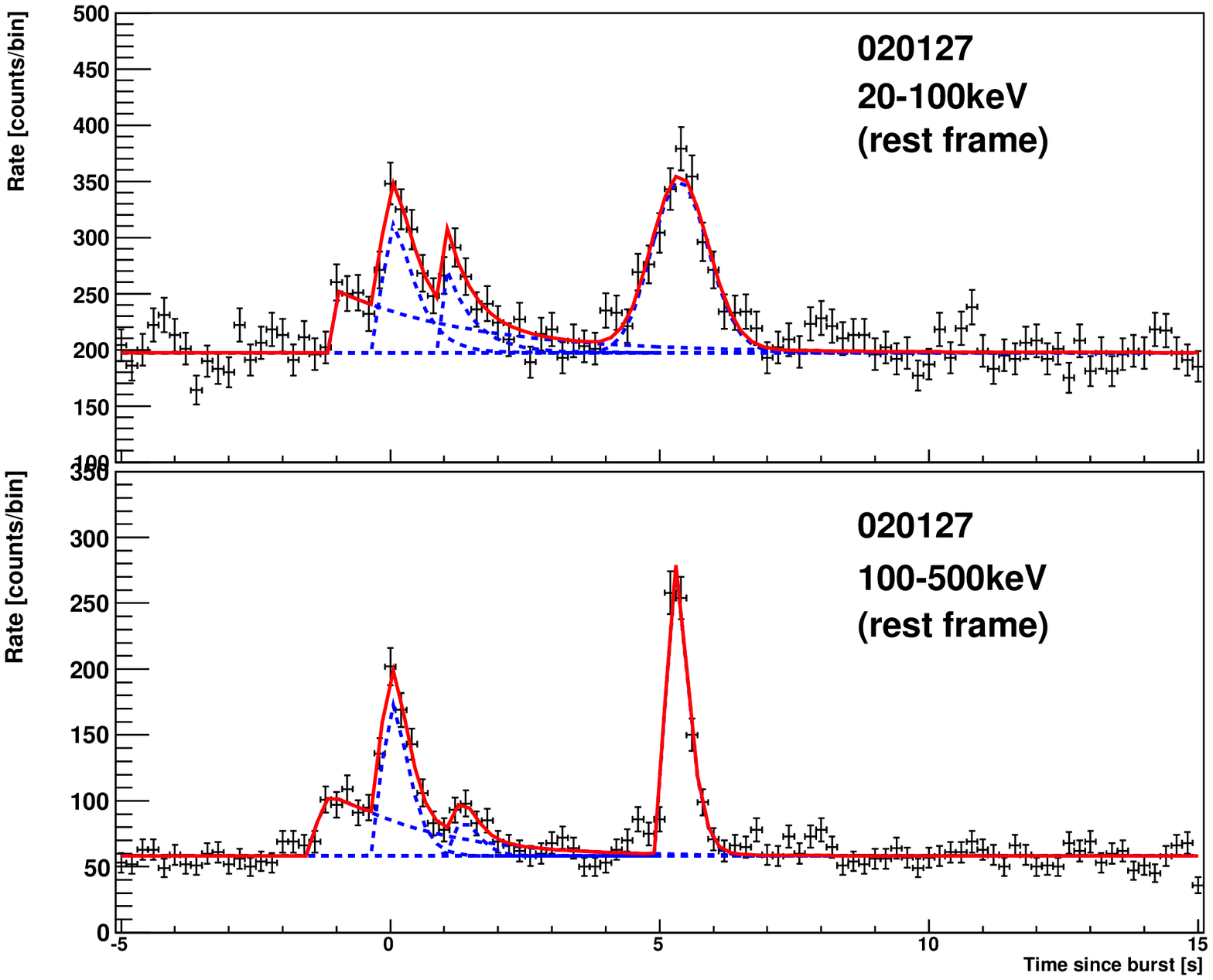}}
  \end{minipage}
  \begin{minipage}[xbt]{75.0mm}
   \resizebox{75.0mm}{!}
   {\includegraphics[angle=0]{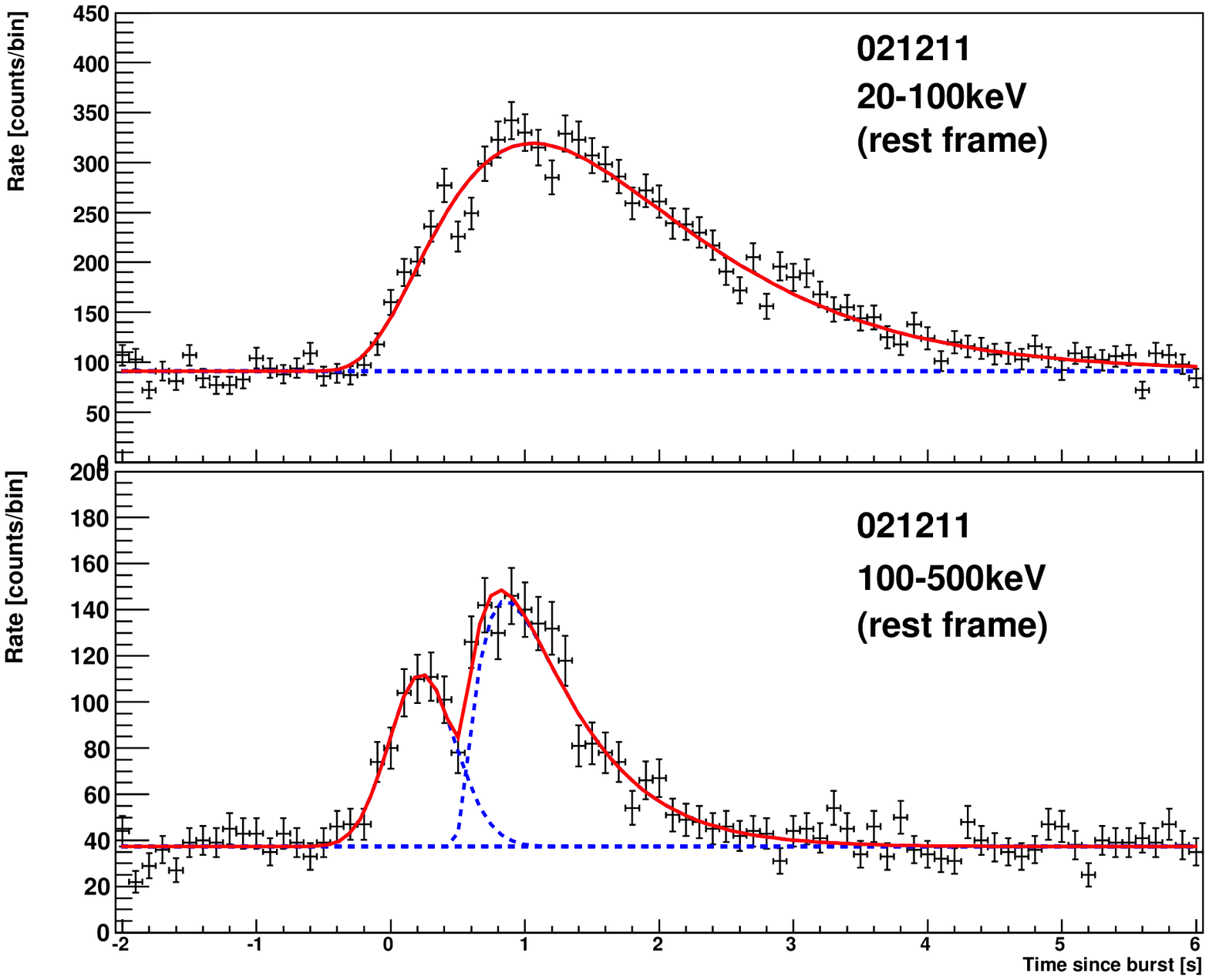}}
  \end{minipage}
 \end{center}
 \begin{center}
  \begin{minipage}[xbt]{75.0mm}
   \resizebox{75.0mm}{!}
   {\includegraphics[angle=0]{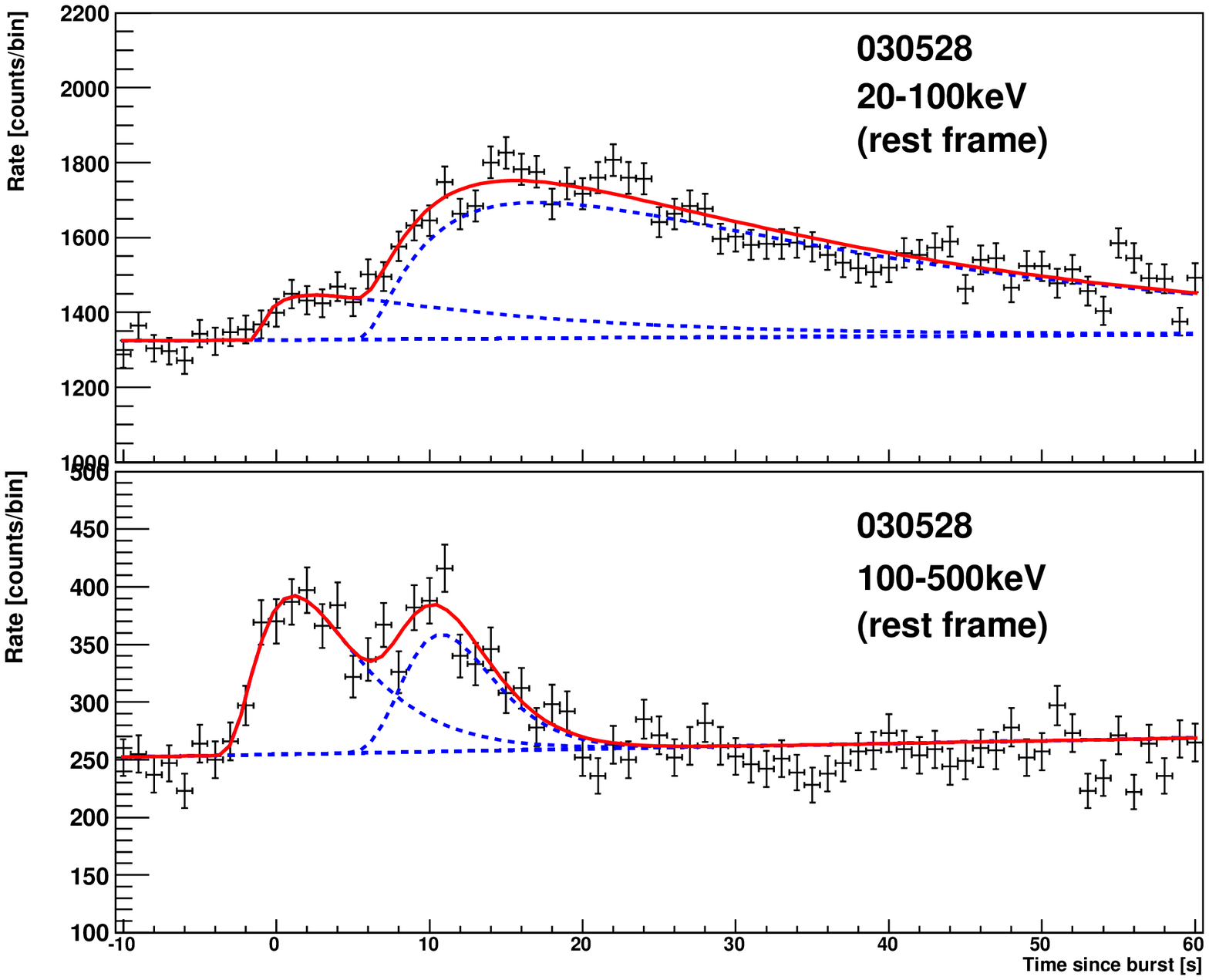}}
  \end{minipage}
  \begin{minipage}[xbt]{75.0mm}
   \resizebox{75.0mm}{!}
   {\includegraphics[angle=0]{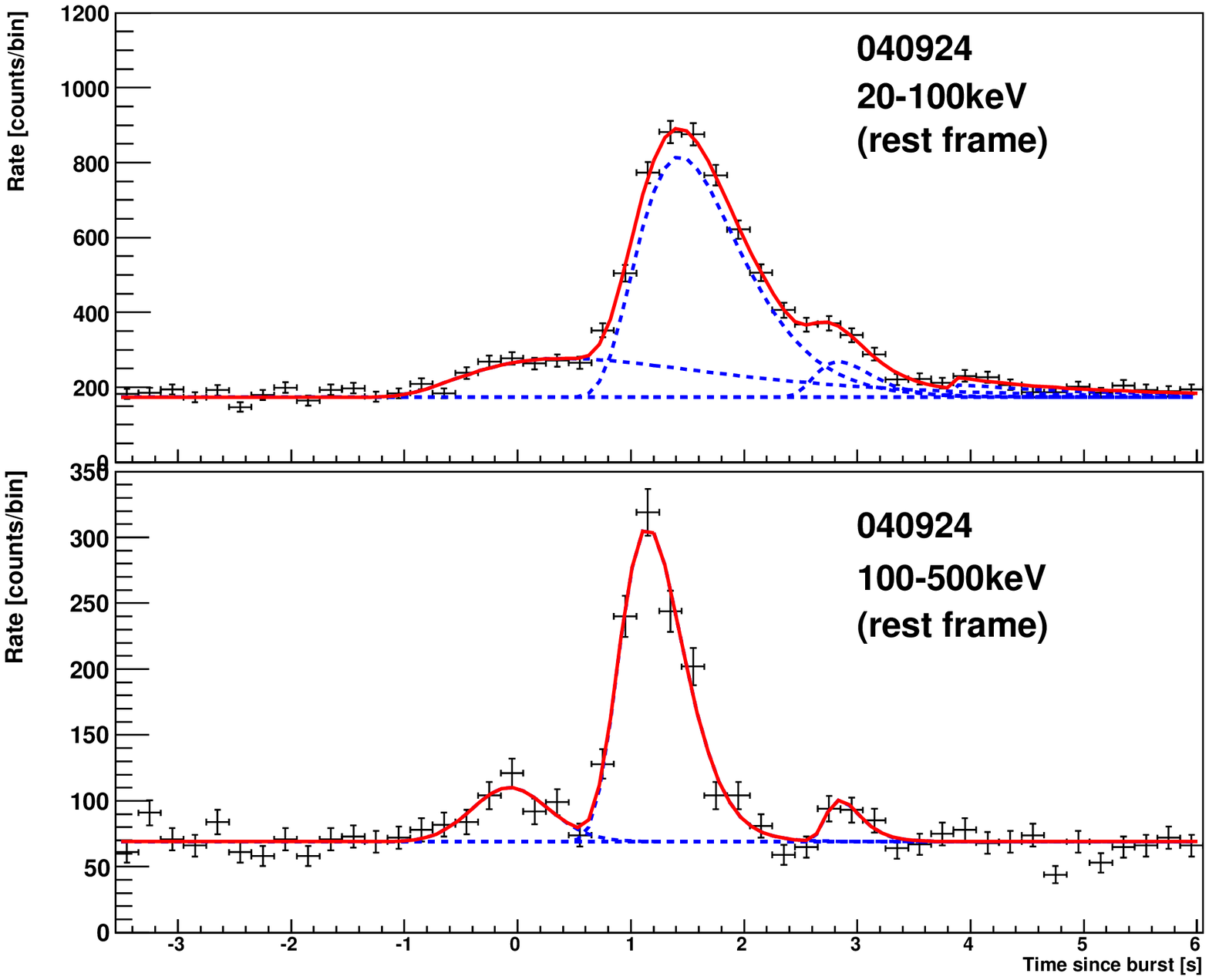}}
  \end{minipage}
 \end{center}
 \begin{center}
  \begin{minipage}[xbt]{75.0mm}
   \resizebox{75.0mm}{!}
   {\includegraphics[angle=0]{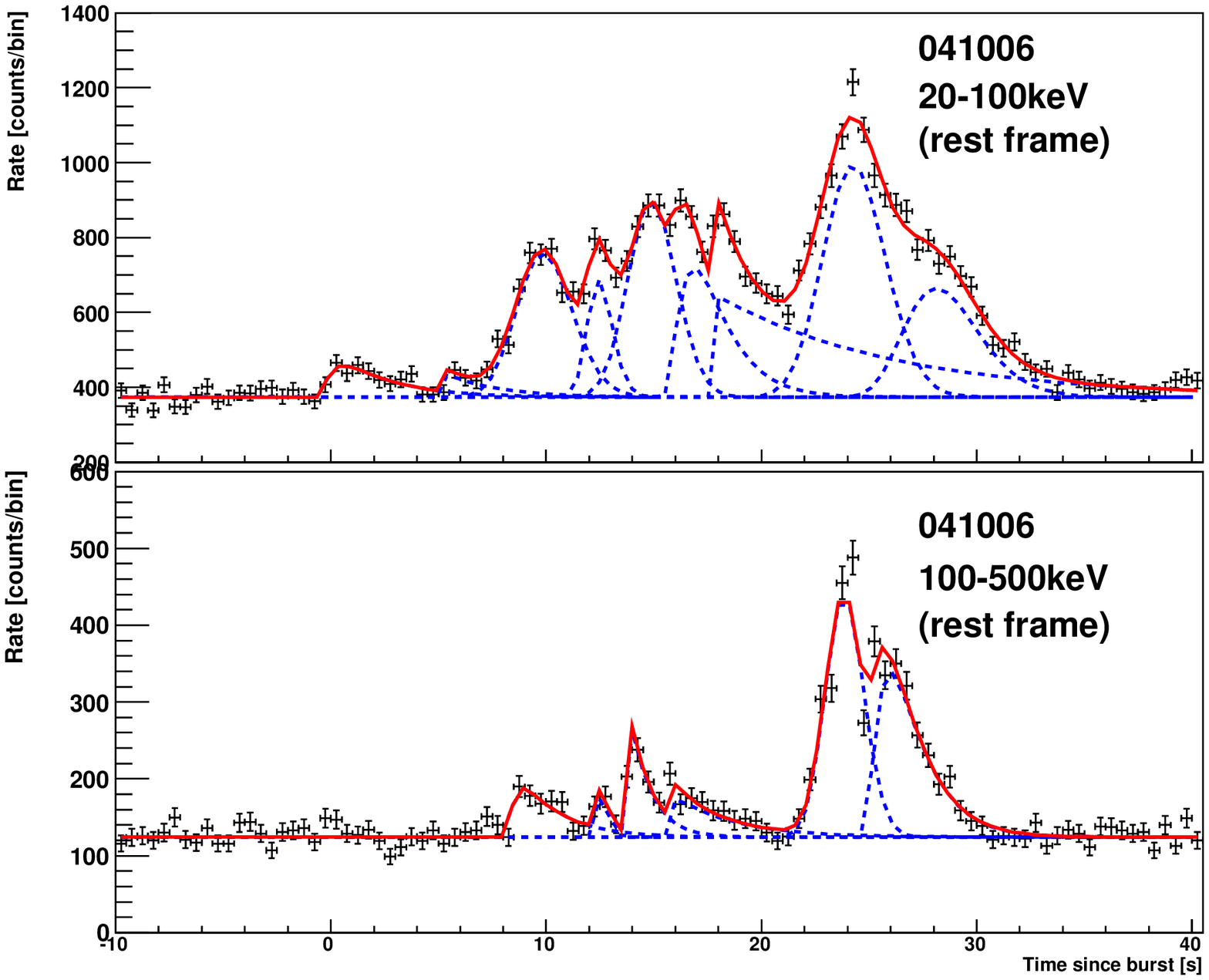}}
  \end{minipage}
  \begin{minipage}[xbt]{75.0mm}
   \resizebox{75.0mm}{!}
   {\includegraphics[angle=0]{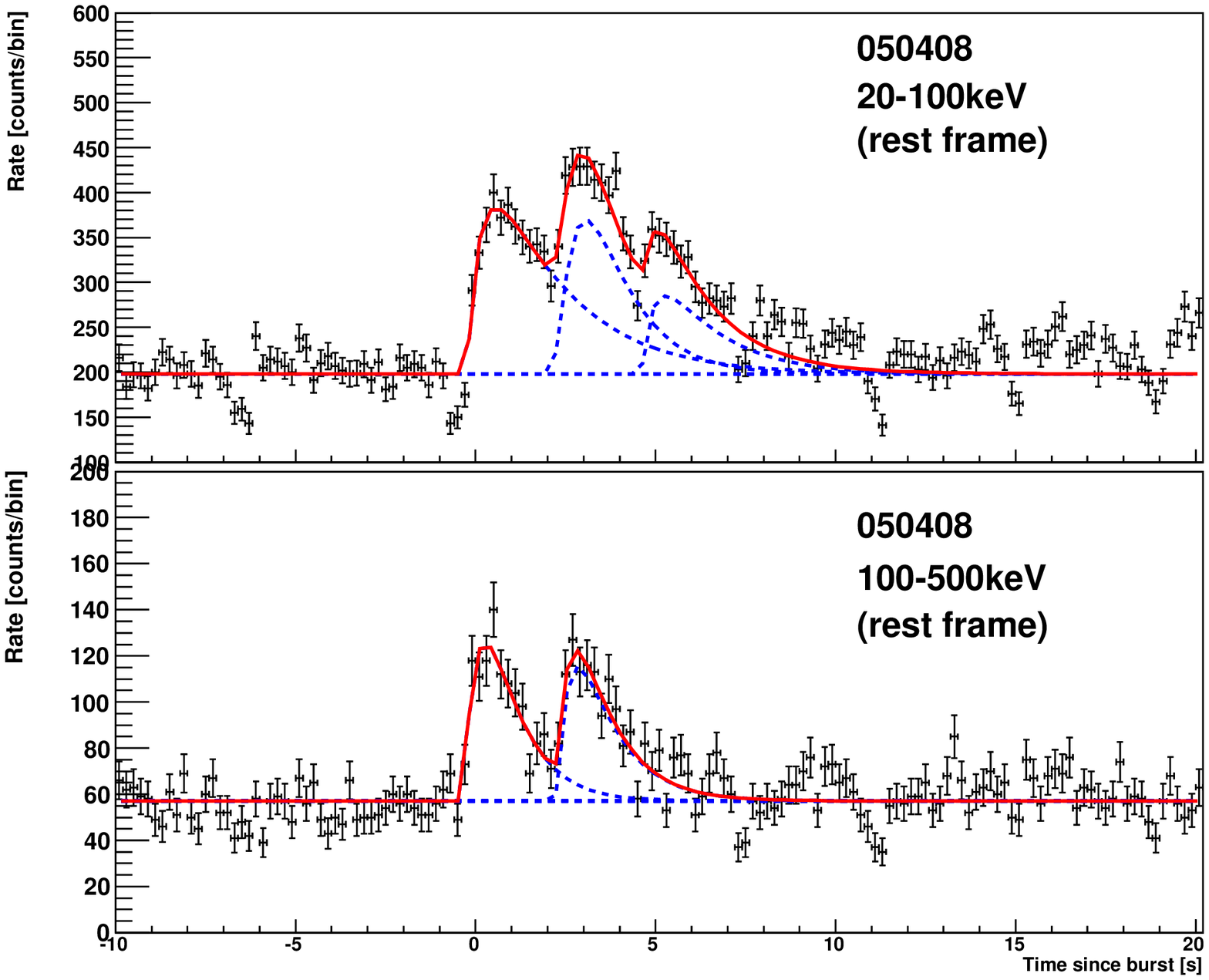}}
  \end{minipage}
  \caption{The result of the fitted pulses in the burst rest frame (20$-$100 keV and 100$-$500 keV).}
  \label{fig:result_fitting_pulse}
 \end{center}
 \end{figure*}

We show the results of the relations between the spectral lag and other 
parameters in the {\it burst rest} frame in Fig. \ref{fig:rest_theoretical_result_lag_lumi_dur} (The result of the fitted pulses is shown in Fig. \ref{fig:result_fitting_pulse}),
 and the adopted energy bands are
determined to have the same energies in common (20$-$100 keV and 100$-$500 keV).
As is the case for the observer's frame, the best-fit parameters of
each relation are summarized in Table \ref{table:param_lag-luminosity}.
We find that there is no significant difference  in results
 between the observer's frame
 and the burst rest frame.
 The obtained results support the idea that the cosmological
effects should not significantly change our measurement in the observer's frame
even though most GRBs are found at high redshifts.
The intrinsic properties for GRB pulses 
predominate over the cosmological effects as suggested by 
\cite{2008ApJ...677L..81H} and \cite{2009AIPC.1133..379H}.

\subsection{Discussion}
We have obtained the correlation between the spectral lag and duration, 
$L_{\rm iso}$ and $E_{\rm peak}$ in the observer's and the burst rest frames. 
In particular, our result extends the energy coverage
to a lower energy band (6$-$25 keV).
This indicates that
the GRB emission in the wide X-ray band has the same origin.

As there is no significant difference between the results in the
observer's and burst rest frames, it is natural to adopt the burst rest
frame to discuss the origin of the spectral lag. Thus, 
in the following discussion, we refer to the case of the burst rest frame.

\subsubsection{Physical Origin of the Relations}
To account for the relation between the spectral lag and $L_{\rm iso}$,
let us consider the off-axis model suggested by
\cite{2001ApJ...554L.163I}; the detector observes
a GRB jet with different viewing angles $\theta_v$. The intrinsic physical parameters,
bulk Lorentz factor $\Gamma$, 
opening half-angle $\Delta\theta \sim 1/\Gamma$, shell radius $r_0$ from the center,
and $E_{\rm peak}^\prime$ in the comoving frame are assumed to be the same for all GRBs.
In this paper, we adopt the same parameters as those of
\cite{2001ApJ...554L.163I}; $\Gamma\Delta\theta$ = 1,
$r_0/c\beta\Gamma^2$ = 1.
This model assumes an intrinsic spectral shape in the comoving frame 
which is 
approximated by the Band function \citep{1993ApJ...413..281B} 
 as
\begin{eqnarray}
f(E^\prime) = \left(\frac{E^\prime}{E_0}\right)^{1+\alpha_B}\left[
 1 + \left(\frac{E^\prime}{E_0}\right)^s \right]^{\frac{\beta_B-\alpha_B}{s}}
\label{eq:Ioka_spectral_shape}
\end{eqnarray}
where $\alpha_B$ and $\beta_B$ are the low- and high-energy indices, 
$s$ describes the smoothness of the transition between the high and low
energies and $E_0$ is the break energy.
\cite{2001ApJ...554L.163I} showed that the observable values such as luminosity,
the pulse duration (FWHM) $w$, and the peak energy $E_{\rm peak}$, change
with the viewing angle $\theta_v$ and correlate with the spectral lag as,
\begin{eqnarray}
L_{\rm iso} &\propto&  \tau_{\rm lag}^{\frac{-2 + \alpha_B}{s+1}},
\label{eq:Ioka_offaxis_model1} \\
w &\propto& 1 + {\rm const} \times \tau_{\rm lag}^\frac{1}{s+1},
\label{eq:Ioka_offaxis_model4} \\
E_{\rm peak} &\propto& \left(1 + {\rm const} \times \tau_{\rm lag}^\frac{1}{s+1}\right)^{-1}
\label{eq:Ioka_offaxis_model5}
\end{eqnarray}

\begin{figure*}[htbp]
 \begin{center}
  \begin{minipage}[xbt]{7.5cm}
   \resizebox{7.5cm}{!}
   {\includegraphics[angle=0]{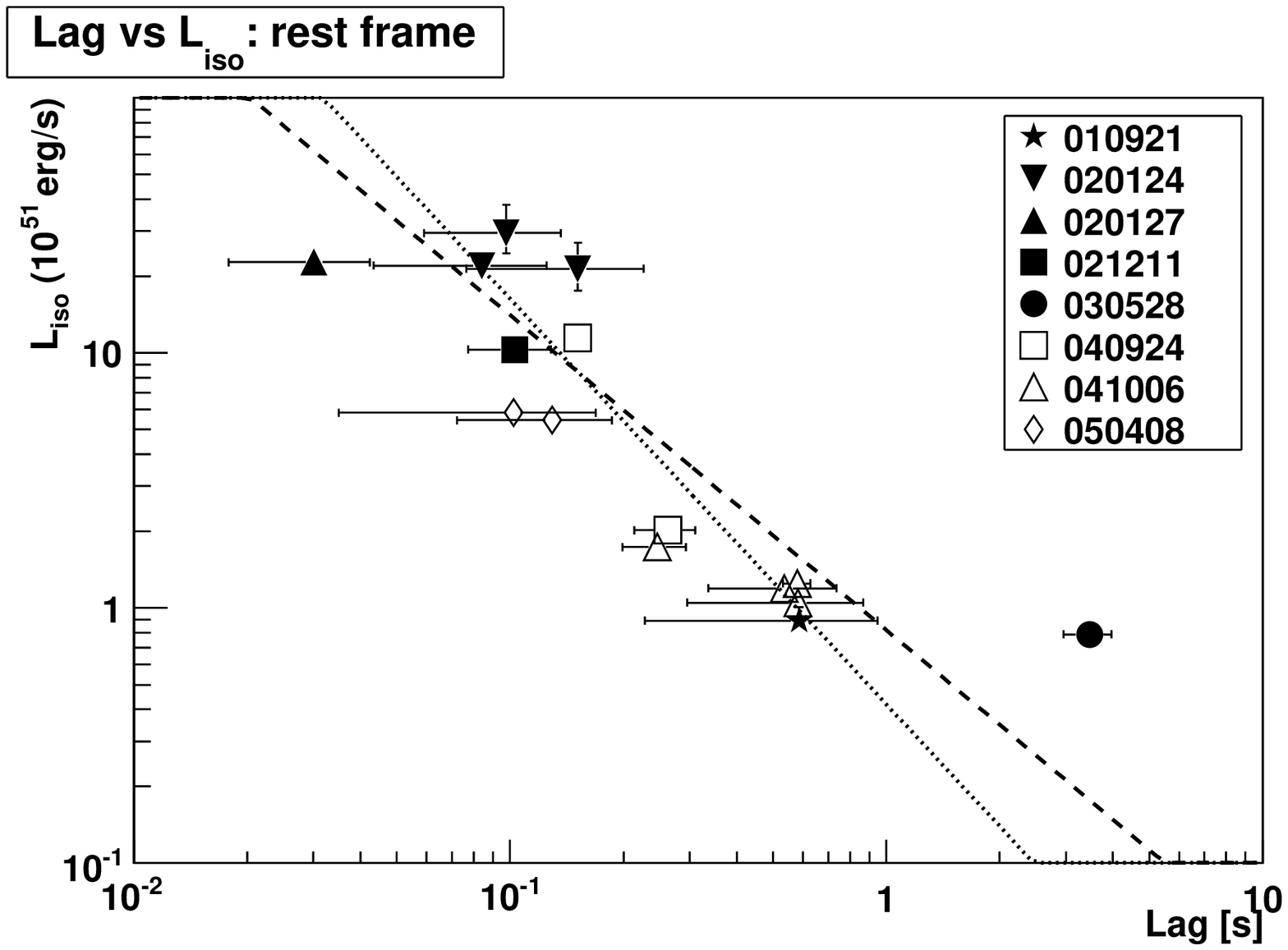}}
  \end{minipage}
  \begin{minipage}[xbt]{7.5cm}
   \resizebox{7.5cm}{!}
   {\includegraphics[angle=0]{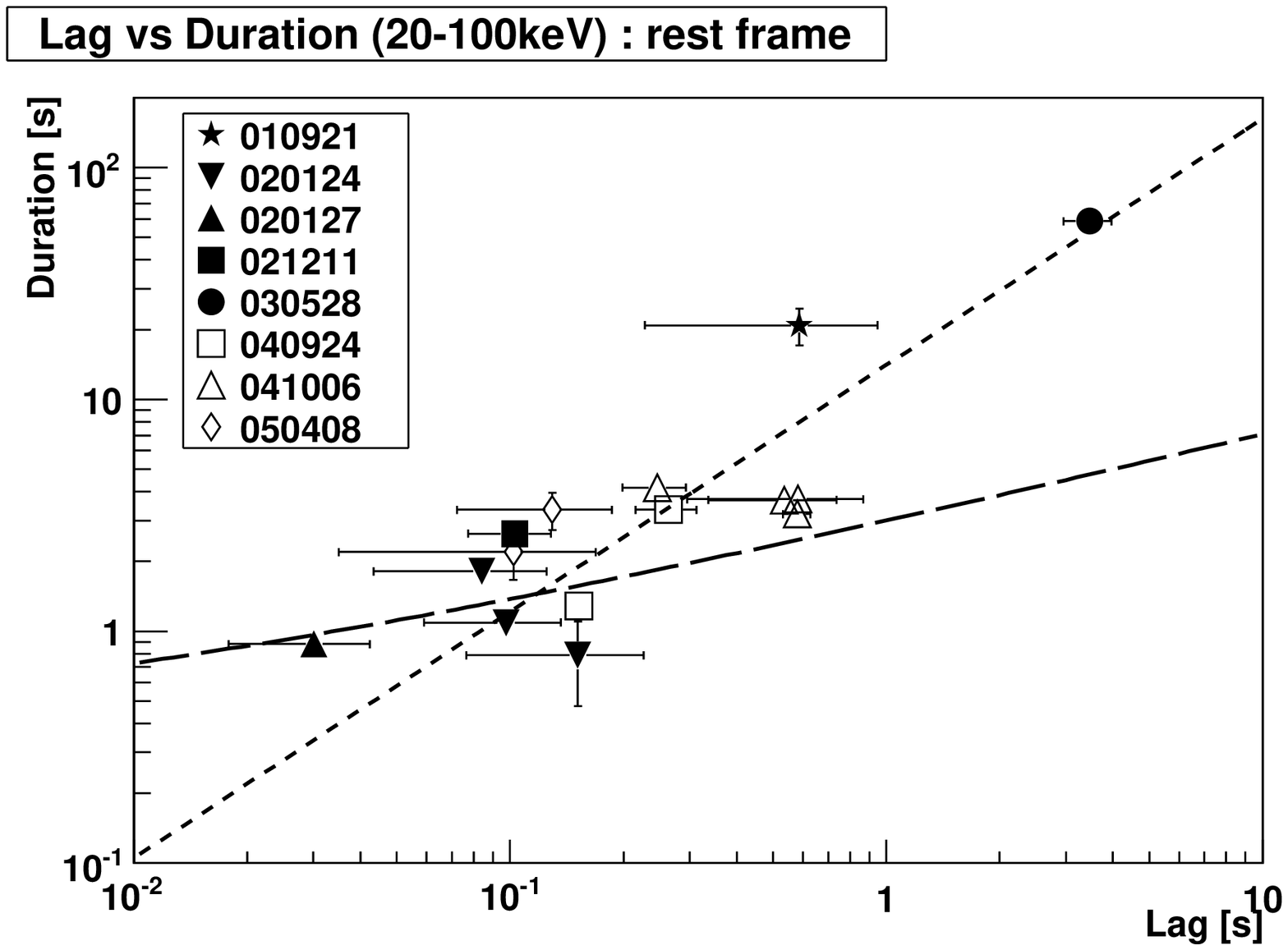}}
  \end{minipage}
  \begin{minipage}[xbt]{7.5cm}
   \resizebox{7.5cm}{!}
   {\includegraphics[angle=0]{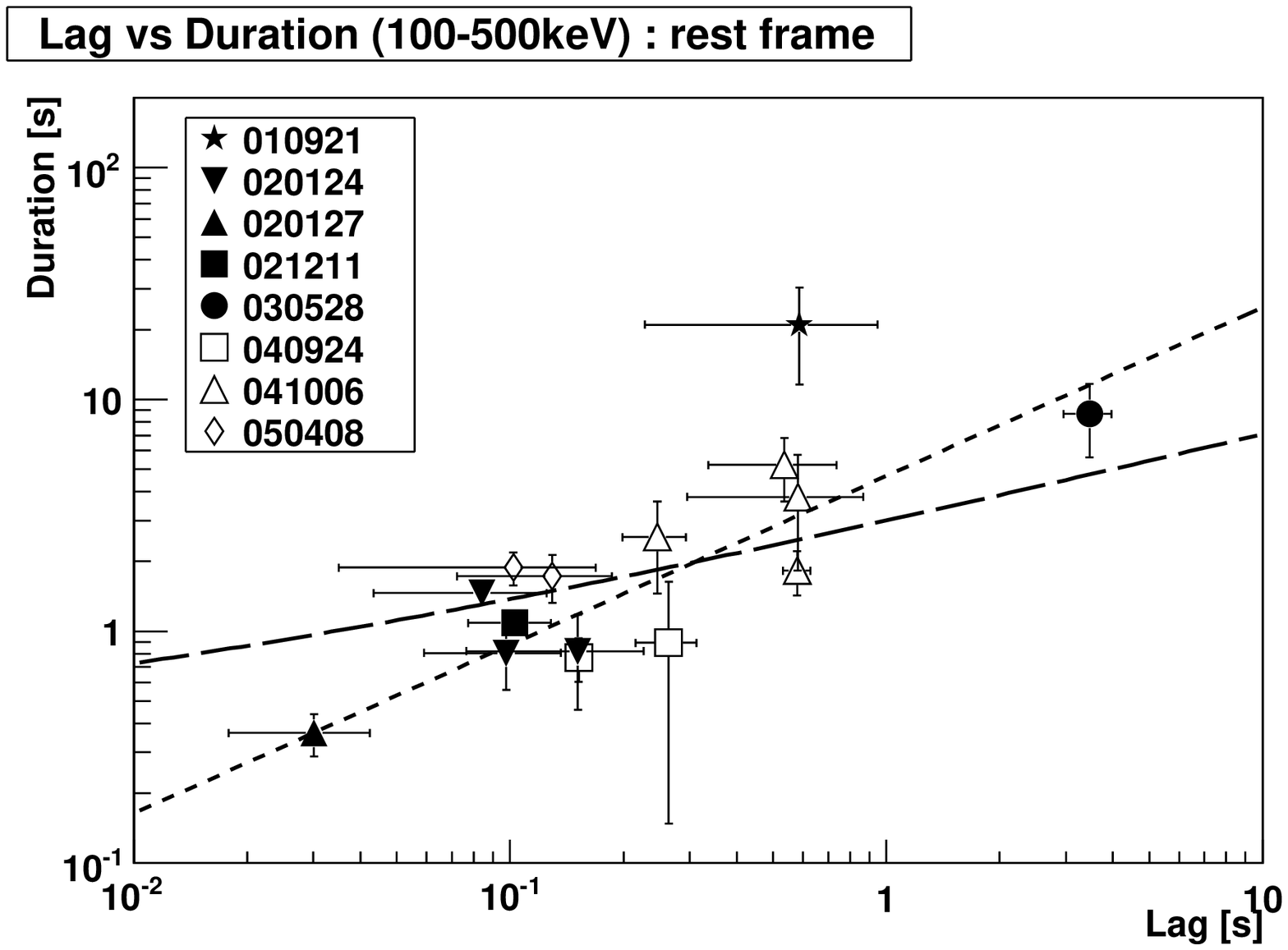}}
  \end{minipage}
  \begin{minipage}[xbt]{7.5cm}
   \resizebox{7.5cm}{!}
   {\includegraphics[angle=0]{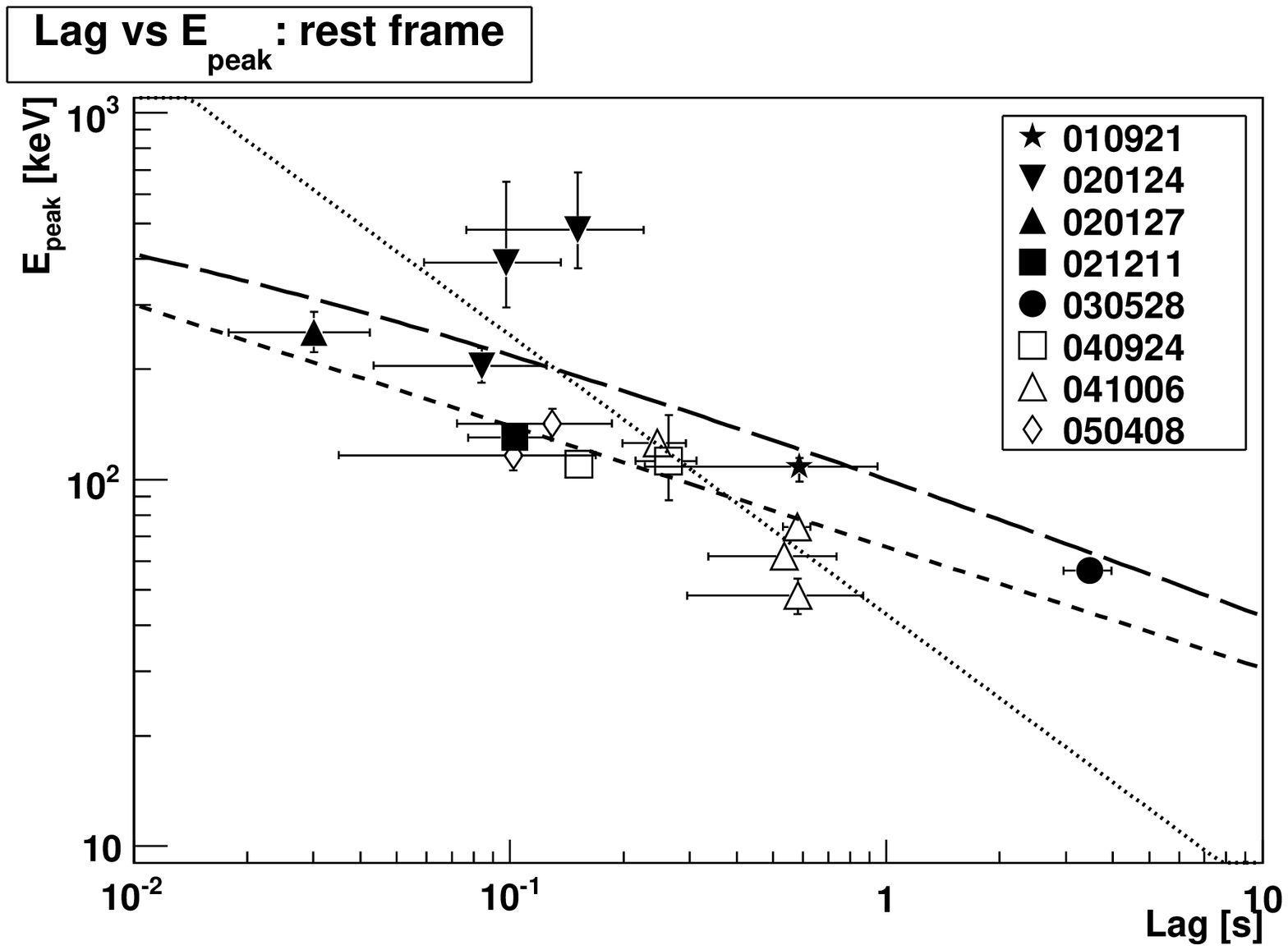}}
  \end{minipage}
 \end{center}
 \caption{Scatter plots ({\it burst rest} frame) of
 spectral lag vs. luminosity (top left), duration for the low-energy band 
(top right), duration for the high-energy band (bottom left), and $E_{\rm peak}$
(bottom right). 
The dashed lines represent the best-fit functions.
For the last three plots, the long dashed lines show the  
  theoretical curve expected in the off-axis model \citep{2001ApJ...554L.163I}. 
The dotted lines represent the best-fit functions including 
the systematic uncertainty and excluding GRB 030528.
 }
  \label{fig:rest_theoretical_result_lag_lumi_dur}
\end{figure*}

We have obtained $\L_{\rm iso} \propto \tau_{\rm lag}^{-1.2}$ from
the lag-luminosity relation in
Fig. \ref{fig:rest_theoretical_result_lag_lumi_dur} so that the off-axis model
with $s = 1.5$ and the typical value for the low-energy photon index $\alpha_B=-1$
can reproduce the lag-luminosity relation.

The expected theoretical results are superimposed on
Fig. \ref{fig:rest_theoretical_result_lag_lumi_dur}. 
Here, we adopt  arbitrary normalization values for the theoretical 
lines. For the lag-duration
relation in the high-energy band (100$-$500 keV), the observational
points are consistent with the theoretical curve. For the lag-duration
relation in the low-energy band (20$-$100 keV), the
observational points agree with the
theoretical curve for small spectral lags, although there are some
outliers for large spectral lags.
For the lag-$E_{\rm peak}$ relation, we find a consistency between the
observational points and the theoretical curve.
Except for some outliers, we find that the off-axis model can explain
the observational results well, even though the model seems to be oversimplified.
(All intrinsic physical parameters are common for all GRBs in this model.)
Furthermore, for the off-axis model the spectral lag is calculated
using the difference between the peak times in the different energy
bands just as we calculate the spectral lag, unlike
\cite{2000ApJ...534..248N}, in which the spectral lags are calculated
using the CCF method.

\subsubsection{Yonetoku Relation}
We now consider the consistency with the Yonetoku relation 
\citep{2004ApJ...609..935Y} in this section.
In the preceding section, we have found that the off-axis model
reproduces the observational results and have not imposed  
any limitations such as the Yonetoku relation ($L_{\rm iso} \propto E_{\rm peak}^{2}$).

Assuming that the Yonetoku relation is valid, from our result on the
lag-luminosity relation ( $L_{\rm iso} \propto \tau_{\rm lag}^{-1.23\pm0.07}$), 
the lag-$E_{\rm peak}$ relation is expected to satisfy 
\begin{eqnarray}
 E_{\rm peak, exp} \propto L_{\rm iso}^{1/2} \propto \tau_{\rm lag}^{-1.23/2}  \propto \tau_{\rm lag}^{-0.62}
\label{eq:Yonetoku_Norris_relation_consistency}
\end{eqnarray}
The index (-0.62) is small compared with the obtained result
($E_{\rm peak, obs} \propto \tau_{\rm lag}^{-0.33\pm0.03})$.
Note that the 
subscripts ``exp'' and ``obs'' represent the expected and observed
values, respectively. 

Let us assume that the determination of the spectral lag has a systematic
uncertainty $\sigma_{\rm sys}$ of 0.05 s resulting from the overlaps of the GRB pulses or
some calibration uncertainties.
In addition we exclude the peculiar case of GRB 030528,
which has a very long spectral lag that may be due to the overlaps of multiple pulses.
Then, the best-fit functions become
$\log(L_{51}) = (-1.38\pm0.13) - (1.59\pm0.28)  \log(\tau_{\rm lag}) $
with reduced chi-square $\chi_\nu^2$ = 8.2/12,
and $\log(E_{\rm peak}) = (1.63\pm0.08) - (0.76\pm0.17)  \log(\tau_{\rm lag}) $
with reduced chi-square $\chi_\nu^2$ = 12.5/12.
The Yonetoku relation and the revised lag-luminosity relation give us
\begin{eqnarray}
 E_{\rm peak, exp} \propto  \tau_{\rm lag}^{-1.59/2}  \propto \tau_{\rm lag}^{-0.80},
\label{eq:Yonetoku_Norris_relation_consistency2}
\end{eqnarray}
which agrees with the revised result (index $-0.76$).
Therefore, considering the small sample and observational uncertainties,
we cannot exclude the validity of the Yonetoku relation from our results.

\section{Detailed Energy Dependence of Spectral Lag and Other Properties}\label{sec:detailed_ene_dep}
Next we consider the detailed energy dependence of the spectral lag and
other properties (the durations including the rise and decay times),
besides the lag-luminosity relation described in the former sections.

\subsection{Energy Dependence of Duration, Rise and Decay Phase}
\cite{2007PASJ...59..857Z} studied the energy dependence of temporal
 properties represented by the formulae,
\begin{eqnarray}
w &\propto& E^{\alpha_{\rm w}}, \\  \tau_{\rm rise} &\equiv& \frac{1}{2}(w-\tau_2)
\propto E^{\alpha_{\rm rise}}, \\  \tau_{\rm decay} &\equiv& \frac{1}{2}(w+\tau_2)
\propto E^{\alpha_{\rm dec}},
\end{eqnarray}
where $\tau_{\rm rise}$ and $\tau_{\rm decay}$
are the rise and decay timescales \citep{2005ApJ...627..324N}.
They found that $\alpha_{\rm w}$ and $\alpha_{\rm dec}$ are highly
correlated, while
$\alpha_{\rm w}$ and $\alpha_{\rm rise}$ are not strongly correlated.
Here, it may be reasonable to assume
that the intrinsic pulse width is responsible for
the rise phase timescale,
while the decay phase timescale is determined by the geometrical
effect in the relativistic expanding shell.
Furthermore, the decay time interval dominates the duration 
because the typical pulse shape shows a fast rise and exponential
decay (FRED).
Thus, it is natural that the decay phase is highly dependent on the
duration and the rise phase is not strongly related to the duration
(or decay time).

\begin{figure}[htbp]
 \begin{center}
  \begin{minipage}[xbt]{95mm}
   \resizebox{95mm}{!}
   {\includegraphics[angle=0]{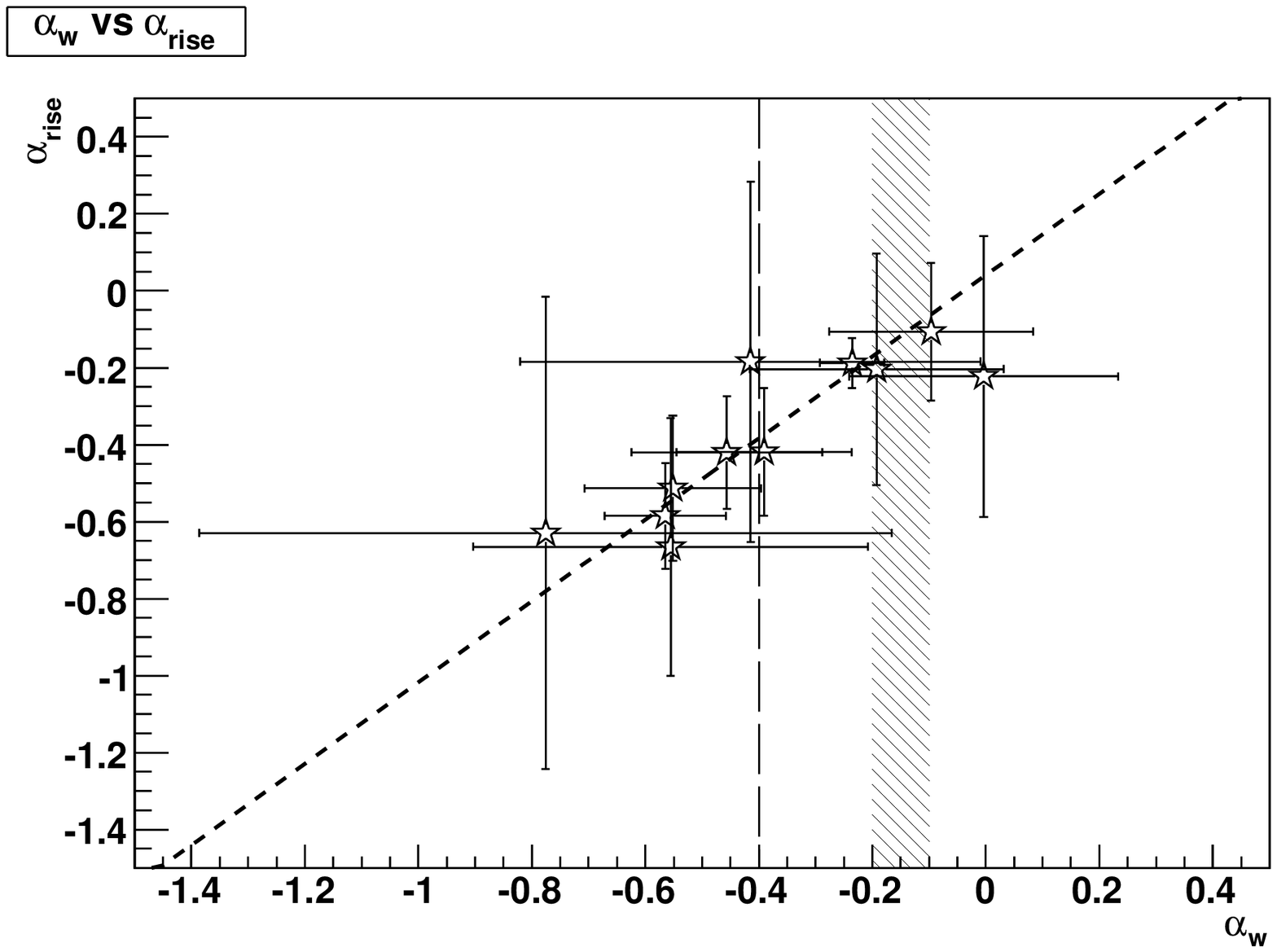}}
  \end{minipage}
  \begin{minipage}[xbt]{95mm}
   \resizebox{95mm}{!}
   {\includegraphics[angle=0]{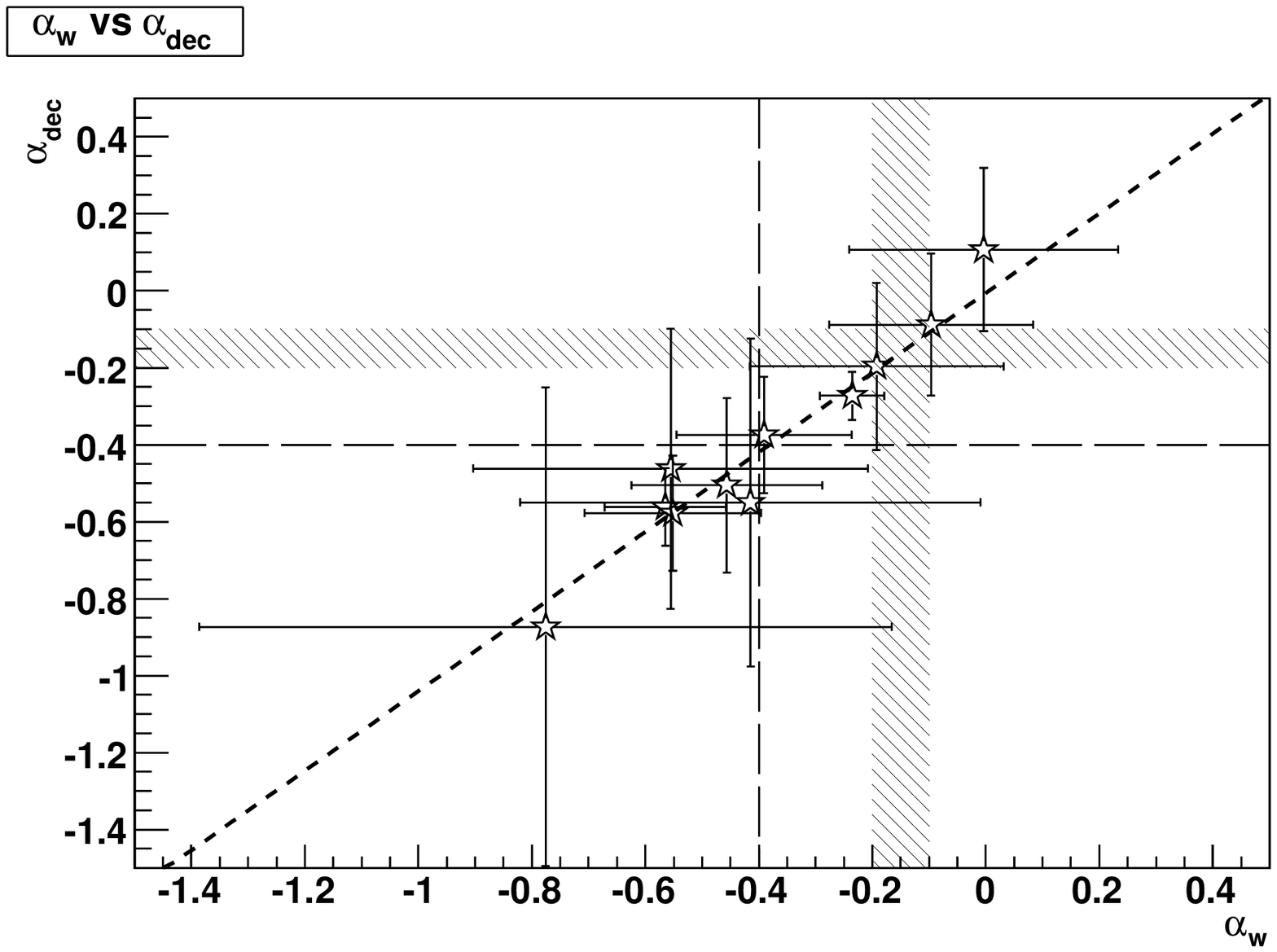}}
  \end{minipage}
 \end{center}
 \begin{center}
  \begin{minipage}[xbt]{95mm}
   \resizebox{95mm}{!}
   {\includegraphics[angle=0]{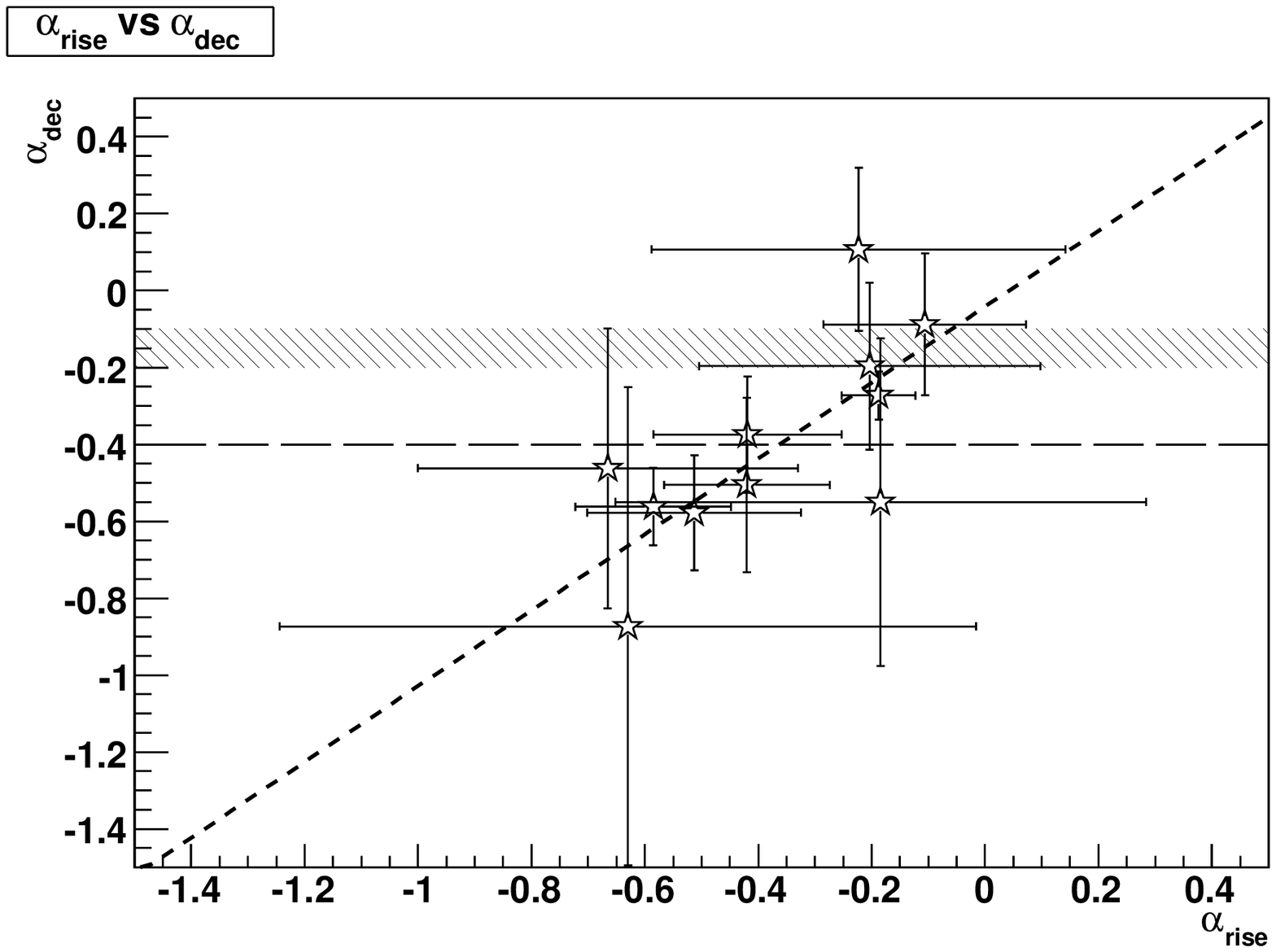}}
  \end{minipage}
  \caption{The
 scatter plots for indices of  $\alpha_{\rm w}$, $\alpha_{\rm  rise}$ and
 $\alpha_{\rm dec}$. 
 ({\it Top}): $\alpha_{\rm w}$ vs. 
  $\alpha_{\rm  rise}$. ({\it Middle}): $\alpha_{\rm w}$ vs. 
  $\alpha_{\rm  dec}$. ({\it Bottom}):  $\alpha_{\rm rise}$ vs. 
  $\alpha_{\rm  dec}$. The dashed line shows the best-fit linear function.
  The
  shaded area and long vertical and horizontal dashed lines represent the expected values from
  the simple curvature \citep{2005MNRAS.362...59S}  and hydrodynamical effects \citep{1998MNRAS.296..275D}, respectively.}
  \label{fig:Scatterplot_FWHM_Rise_Dec}
 \end{center}
\end{figure}

\begin{figure*}[htbp]
 \begin{center}
  \begin{minipage}[xbt]{50.0mm}
   \resizebox{50.0mm}{!}
   {\includegraphics[angle=0]{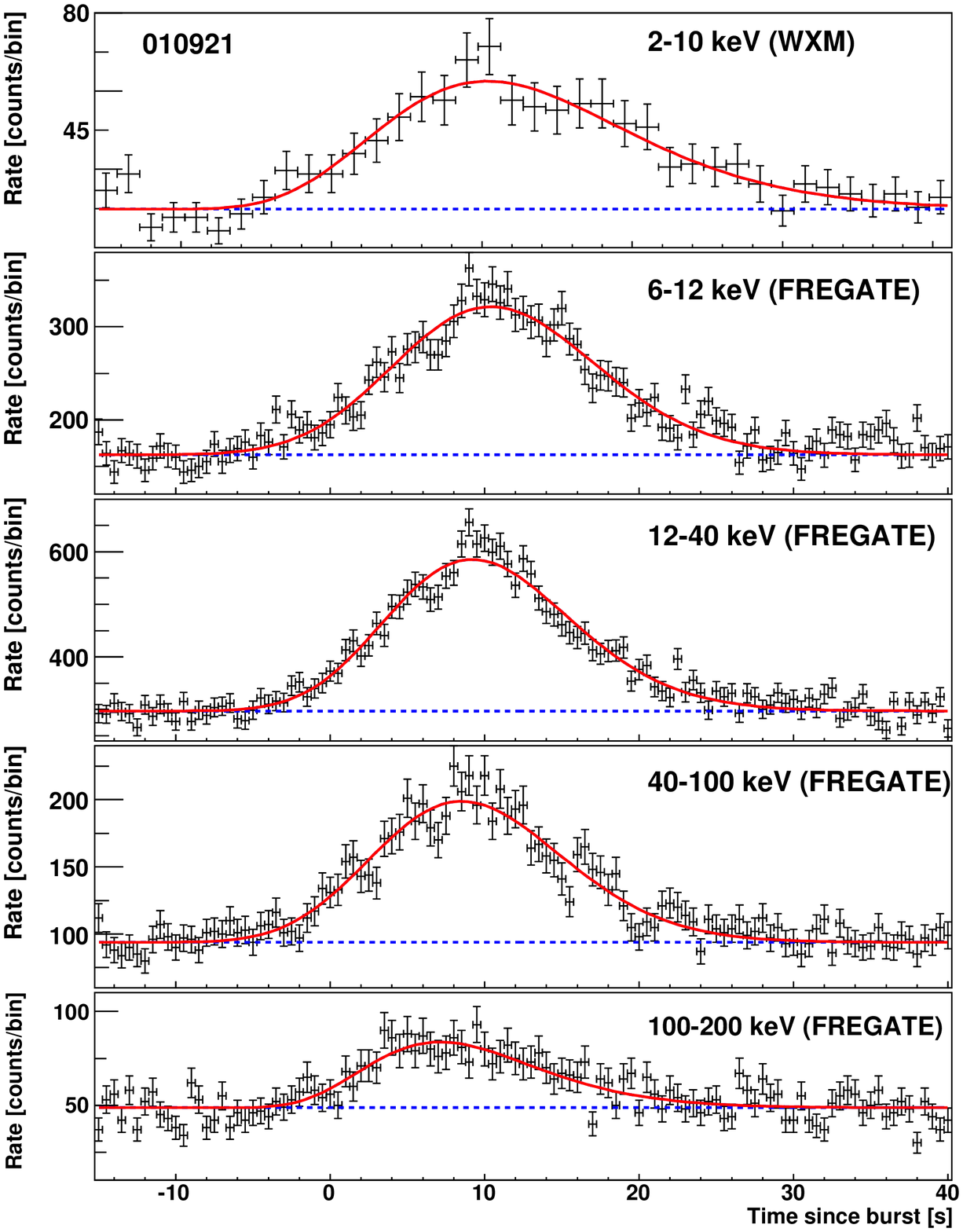}}
  \end{minipage}
  \begin{minipage}[xbt]{50.0mm}
   \resizebox{50.0mm}{!}
   {\includegraphics[angle=0]{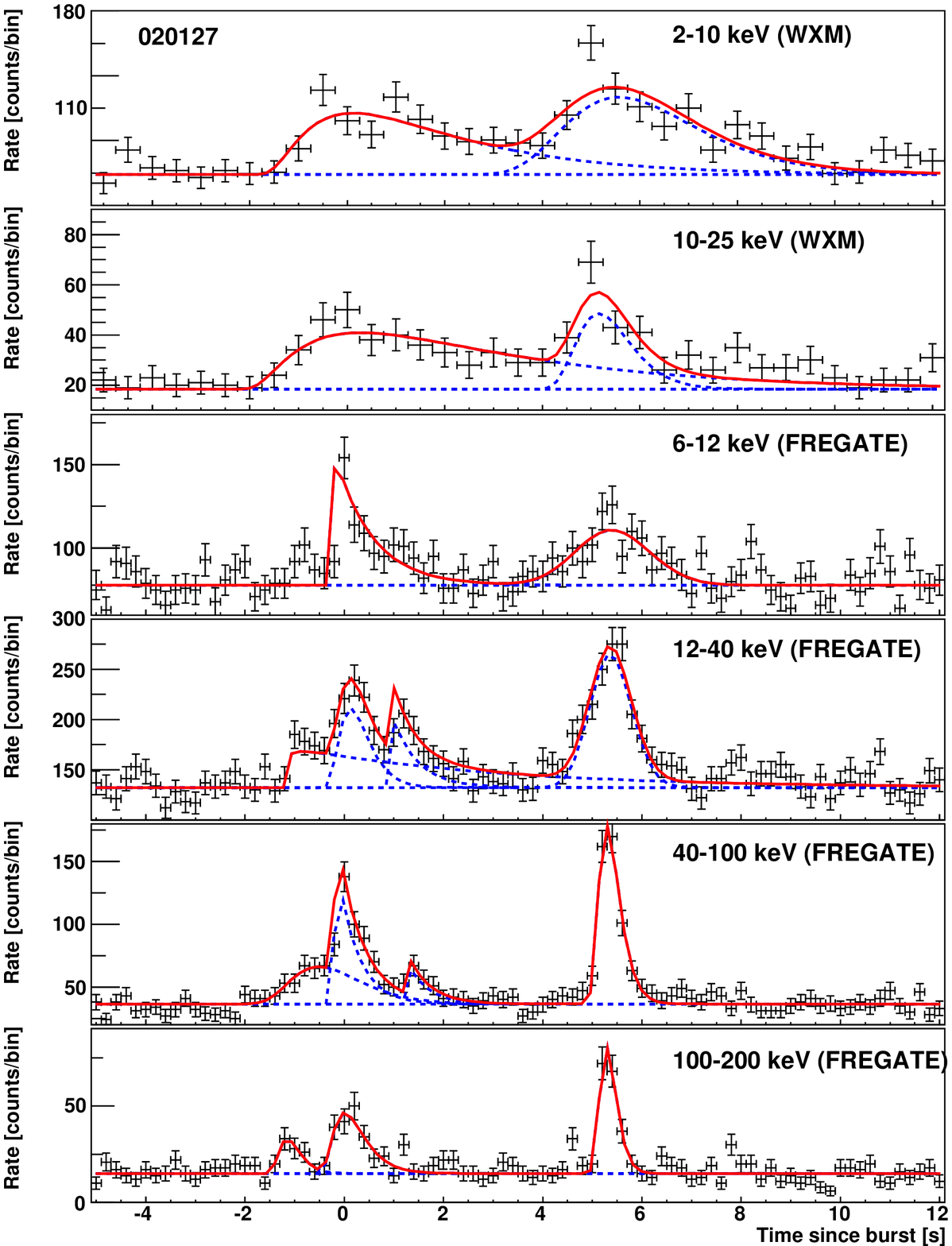}}
  \end{minipage}
  \begin{minipage}[xbt]{50.0mm}
   \resizebox{50.0mm}{!}
   {\includegraphics[angle=0]{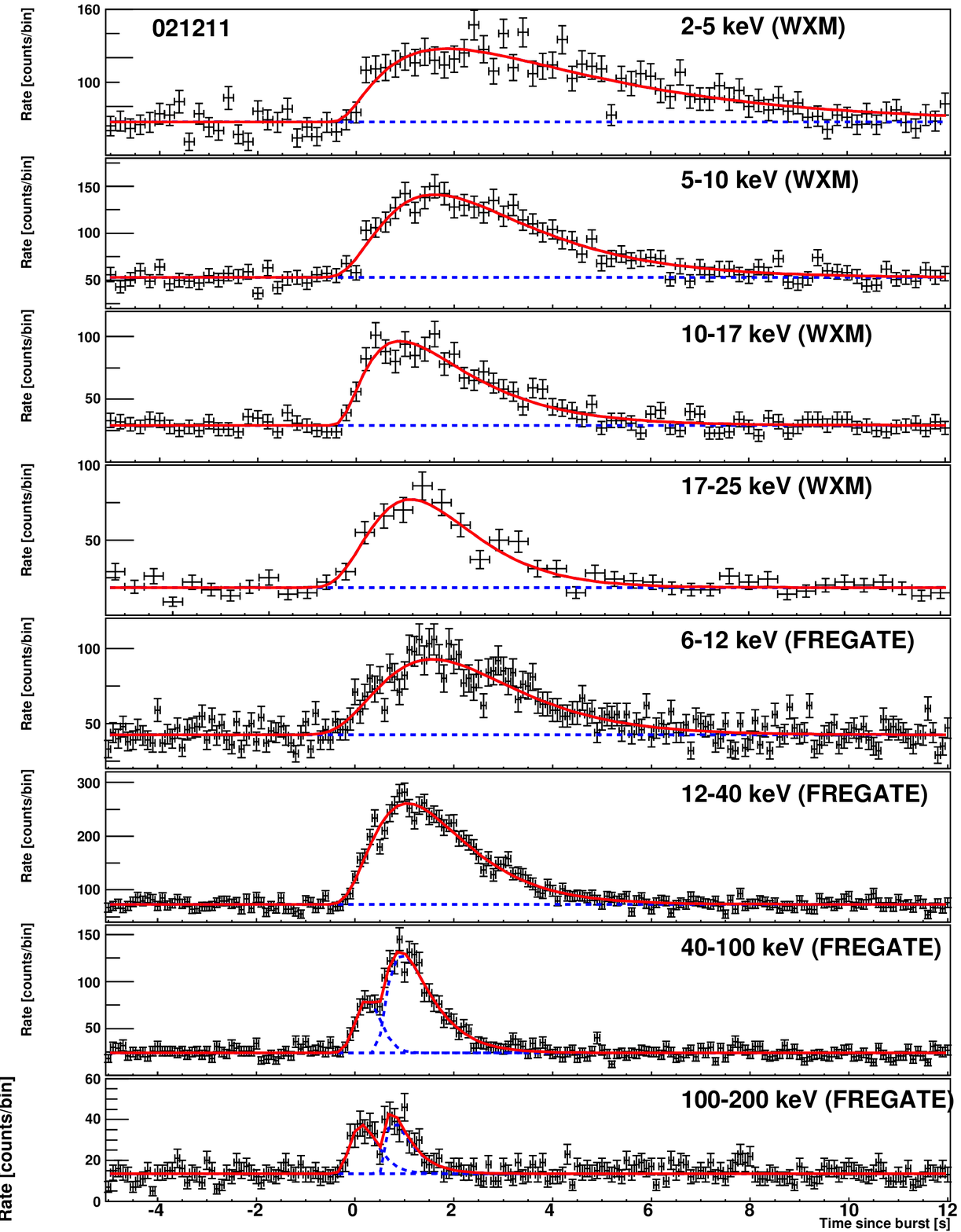}}
  \end{minipage}
 \end{center}
 \begin{center}
  \begin{minipage}[xbt]{50.0mm}
   \resizebox{50.0mm}{!}
   {\includegraphics[angle=0]{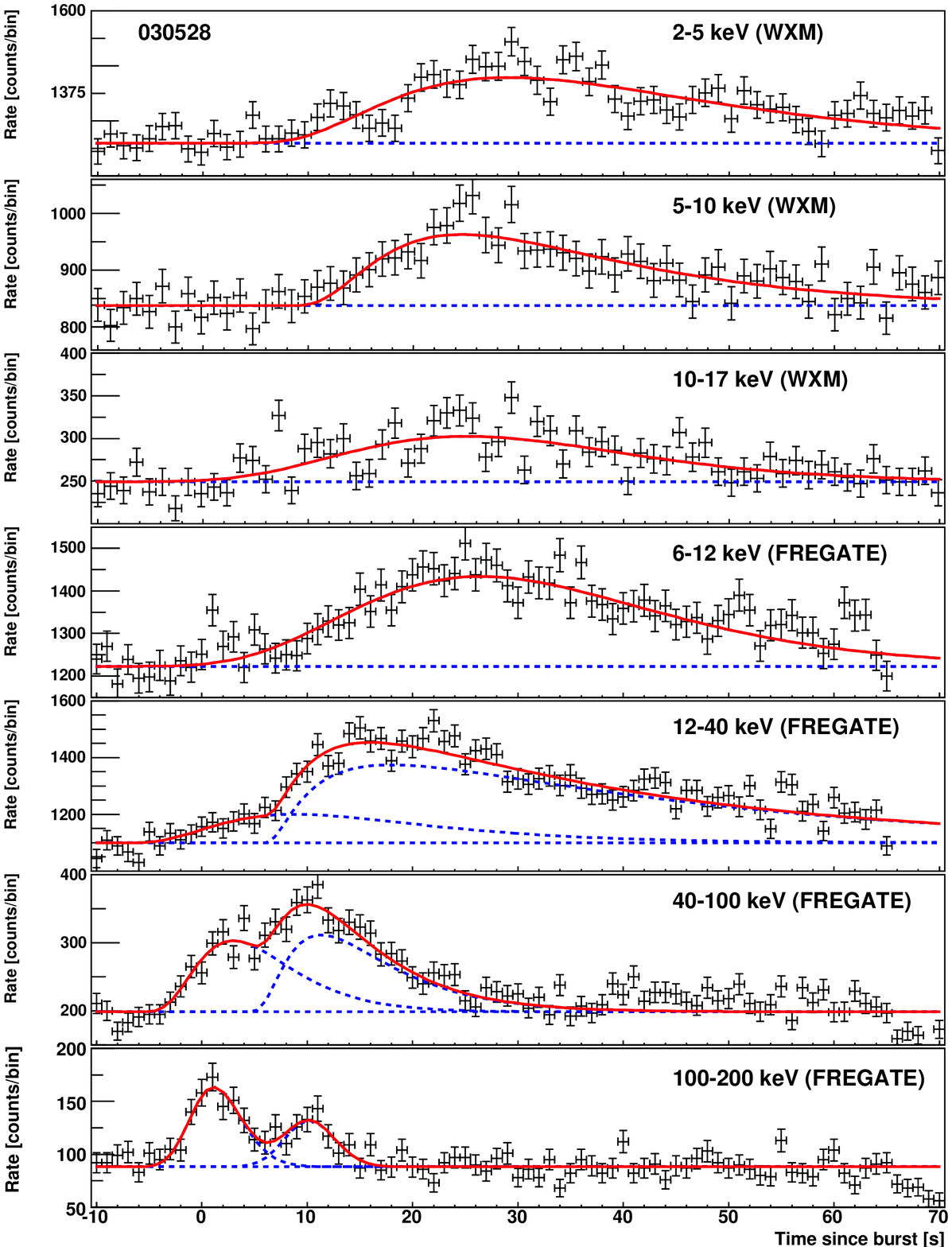}}
  \end{minipage}
  \begin{minipage}[xbt]{50.0mm}
   \resizebox{50.0mm}{!}
   {\includegraphics[angle=0]{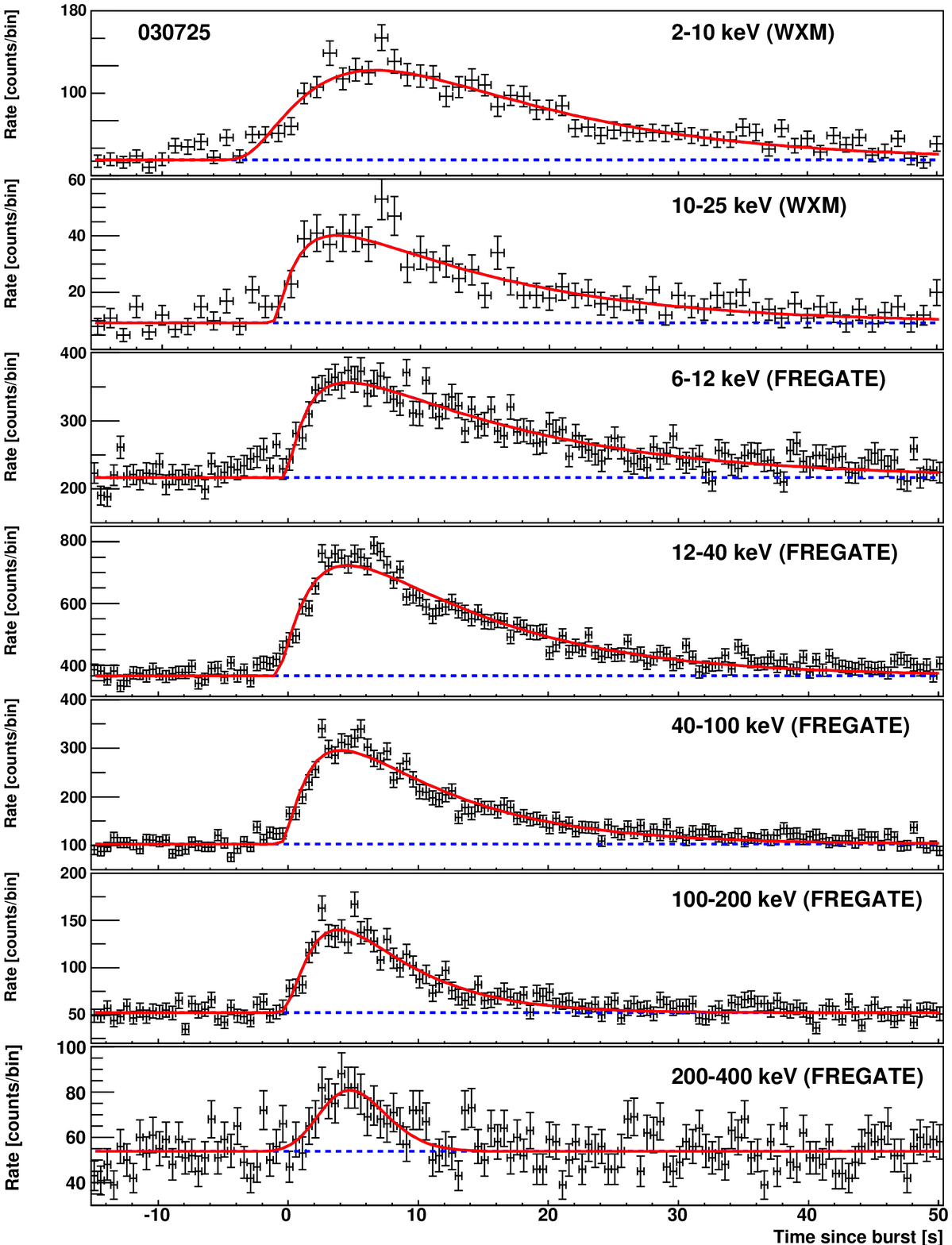}}
  \end{minipage}
  \begin{minipage}[xbt]{50.0mm}
   \resizebox{50.0mm}{!}
   {\includegraphics[angle=0]{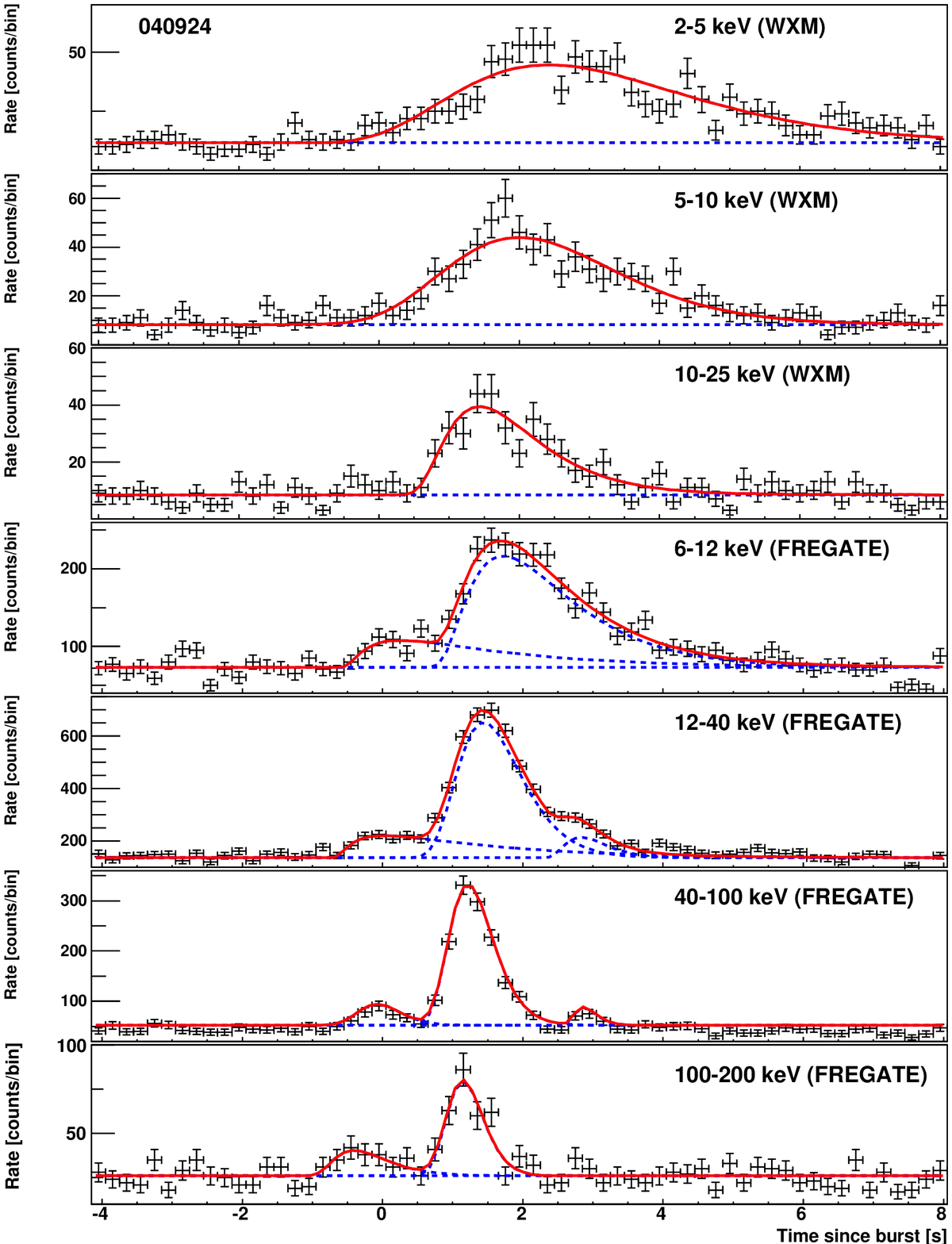}}
  \end{minipage}
 \end{center}
 \begin{center}
  \begin{minipage}[xbt]{50.0mm}
   \resizebox{50.0mm}{!}
   {\includegraphics[angle=0]{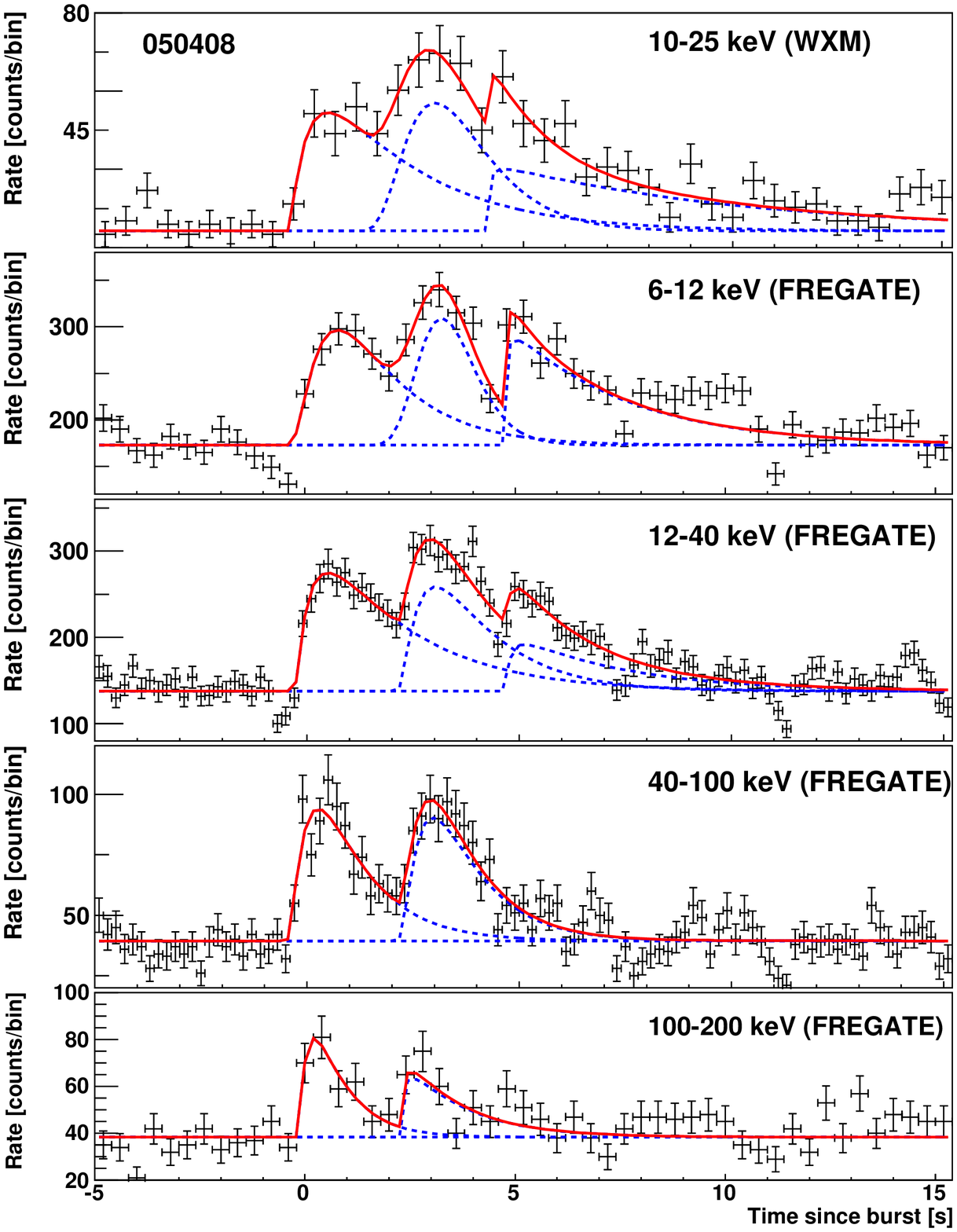}}
  \end{minipage}
  \begin{minipage}[xbt]{50.0mm}
   \resizebox{50.0mm}{!}
   {\includegraphics[angle=0]{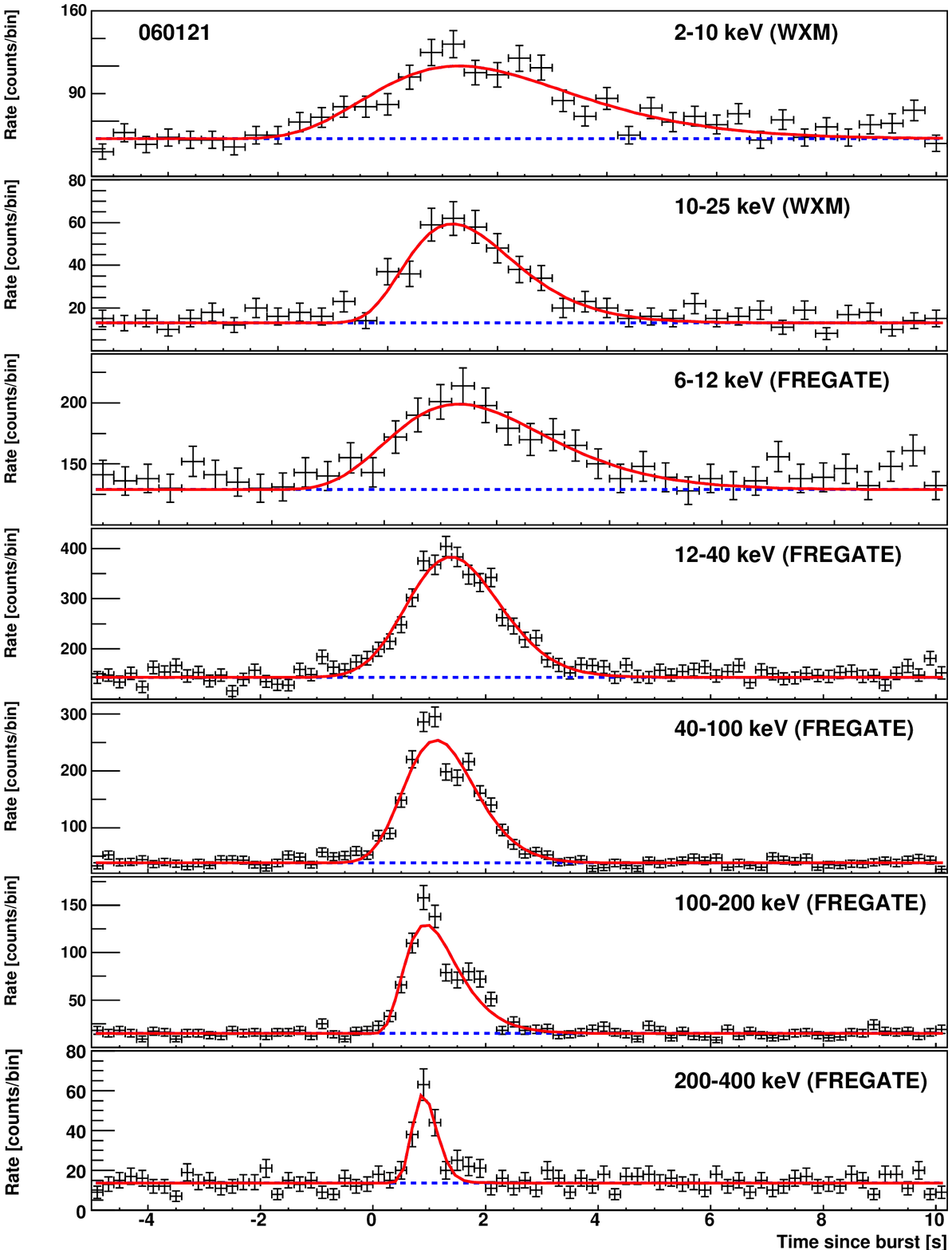}}
  \end{minipage}
  \caption{The result of fitted pulses in multiple energy bands.}
  \label{fig:result_fitting_pulse_multi}
 \end{center}
 \end{figure*}

We show the results of the plots among $\alpha_{\rm w}$, $\alpha_{\rm rise}$
and $\alpha_{\rm dec}$ in Fig. \ref{fig:Scatterplot_FWHM_Rise_Dec} (The result of the fitted pulses is shown in Fig. \ref{fig:result_fitting_pulse_multi}).
The top panel of Fig. \ref{fig:Scatterplot_FWHM_Rise_Dec} shows the
scatter plot of $\alpha_{\rm w}$ versus $\alpha_{\rm rise}$ in our
{\it HETE-2} sample. Although
the uncertainty for each point is very large, we find a marginal
linear relation with correlation coefficient $R = 0.51^{+0.18}_{-0.38}$.
The best-fit function is 
$\alpha_{\rm rise} = (0.04 \pm 0.15) + (1.06 \pm 0.41)\alpha_{\rm w}$.
If the rise timescale is determined only by the intrinsic pulse width,
the correlation between $\alpha_{\rm rise}$ and $\alpha_{\rm w}$ should be weak.
However, from this result 
$\alpha_{\rm w}$ seems to be roughly proportional to 
$\alpha_{\rm rise}$.
Therefore, the rise timescale depends not only on the
the intrinsic pulse width but also somewhat on the geometrical (curvature) effect.
 Some previous studies may give us clues to understanding
this behavior;
\cite{2007ApJ...663.1110L} and \cite{2009ApJ...698..417P} found that 
$E_{\rm peak}$ decays monotonically through long GRB pulses.
This energy decay occurs even prior to the pulse peak,
namely in the rise time phase.
Therefore, the pulse rise phase
is a part of the $E_{\rm peak}$-decay phase.
\cite{2009AIPC.1133..379H} also demonstrated that 
the high-energy pulse intensity is starting to decline prior to the pulse peak 
in the low-energy band, as is the case for the {\it HETE-2} results.
Thus, these results indirectly indicate
that the pulse rise timescale is affected by the 
pulse decay time.

The middle panel of Fig. \ref{fig:Scatterplot_FWHM_Rise_Dec} shows the
scatter plot of $\alpha_{\rm w}$ versus $\alpha_{\rm dec}$.
The best-fit function is 
$\alpha_{\rm dec} = (-0.01 \pm 0.13) + (1.03 \pm 0.37)\alpha_{\rm w}$
with correlation coefficient $R = 0.67^{+0.13}_{-0.37} $.
 Since a relatively good proportionality between $\alpha_{\rm w}$ and
 $\alpha_{\rm dec}$ exists,  this result 
supports the approximation $w \sim \tau_{\rm decay}$ and
the assumption that the curvature effect determines the
decay timescale.

The relations between $\alpha_{\rm rise}$ versus $\alpha_{\rm dec}$ are plotted
in the bottom panel of Fig. \ref{fig:Scatterplot_FWHM_Rise_Dec},
where the best-fit function is 
$\alpha_{\rm dec} = (-0.04 \pm 0.13) + (0.99 \pm 0.40)\alpha_{\rm rise}$
with correlation coefficient $R = 0.46^{+0.20}_{-0.38} $.
The result also seems to show that the rise timescale is slightly affected
by the curvature effect.
Although the uncertainties in the correlation coefficient and fitting
parameters are large, the
relationships between $\alpha_{\rm w}$,  $\alpha_{\rm rise}$ and
$\alpha_{\rm dec}$ are consistent with those of 
 \cite{2007PASJ...59..857Z} (the three functional forms agree with ours).

\cite{2005MNRAS.362...59S} computed the temporal profiles
of the GRB pulse in the four BATSE energy bands, with the relativistic
curvature effect of an expanding shell.
They included an intrinsic ``Band'' shape spectrum and an intrinsic
energy-independent emission profile, and estimated the dependence
of the duration and other properties on energy as
$w$ $\propto$ $E^{-0.2 \sim -0.1}$ 
($\alpha_{\rm w} \simeq \alpha_{\rm dec} = -0.2$ to $-0.1$).
On the other hand, \cite{1998MNRAS.296..275D} calculated the time-evolution
of the internal shocks (hydrodynamical effect)
assuming a  highly non-uniform distribution of the Lorentz factor,
and obtained the energy dependence as
$w \propto$ $E^{-0.4}$.

In our result, shown in Fig. \ref{fig:Scatterplot_FWHM_Rise_Dec}, 
$\alpha_{\rm w}$ and $\alpha_{\rm dec}$ range from -0.8 to 0 and 
the expected energy dependences for the models of \cite{1998MNRAS.296..275D}
and \cite{2005MNRAS.362...59S} are represented as a long dashed line and a shaded
portion, respectively.
The data points are  widely scattered
so that the simple model of \cite{1998MNRAS.296..275D} 
or \cite{2005MNRAS.362...59S} alone cannot explain the results we obtained.

\subsection{Physical Origin of the Spectral Lag for Individual Pulses}
In this section, we try to clarify the origin of the spectral lag of our 
{\it HETE-2} GRBs, apart from the lag-luminosity relation described in
the preceding section.
First we need to check whether only the simple curvature effect,
which should be in any case included, can explain the energy dependence of the lag or not.
Here we consider the curvature effect model
described by \cite{2006MNRAS.367..275L}. They
calculated the light curves from
an isotropically expanding sphere with a constant bulk Lorentz factor
and the Band function for a rest-frame radiation spectrum.
A Gaussian pulse was assumed for the light curve in the source rest frame.
While the generic formula for the spectral lag is complicated,
they demonstrated that the spectral 
lag has a energy dependence with lag $\propto$ $E$ below a saturated energy,
$E_s=1.67 E_{\rm peak}$ for the typical parameter sets of the low-energy index
$\alpha_B$ = -1, high-energy index $\beta_B$ = -2.25 and the shell radius $r_0$
= 3 $\times$ 10$^{15}$ cm.
Considering the beaming effect, photons of
energy at $E > E_s$ would mainly come from the area of the
surface around the line of sight, i.e., $\theta \lesssim \Gamma^{-1}$ (where
$\theta$ is the angle to the line of sight).
When $E > E_s$, the contribution to the corresponding light
curve largely comes from the high-energy portion of the rest frame
spectrum, which causes the peak time of the light curve to change less, and
the lag would saturate. 
On the other hand, for $E < E_{\rm s}$, ``off-axis'' ($\theta>\Gamma^{-1}$)
photons may contribute to the light curve so that
the lag increases with the increasing energy difference in two energy bands.

\begin{figure*}[htbp]
 \begin{center}
  \begin{minipage}[xbt]{75mm}
   \resizebox{75mm}{!}
   {\includegraphics[angle=0]{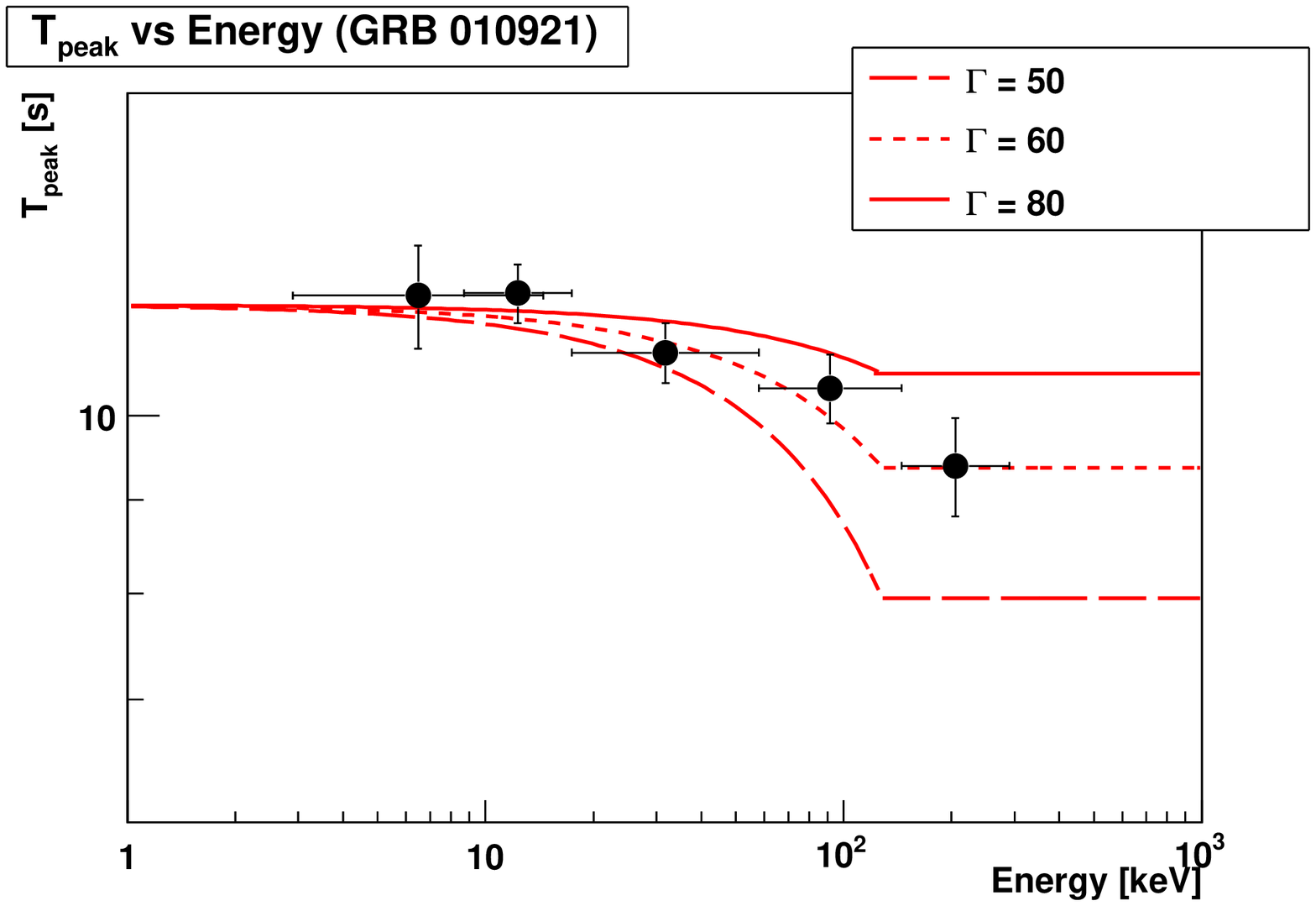}}
  \end{minipage}
  \begin{minipage}[xbt]{75mm}
   \resizebox{75mm}{!}
   {\includegraphics[angle=0]{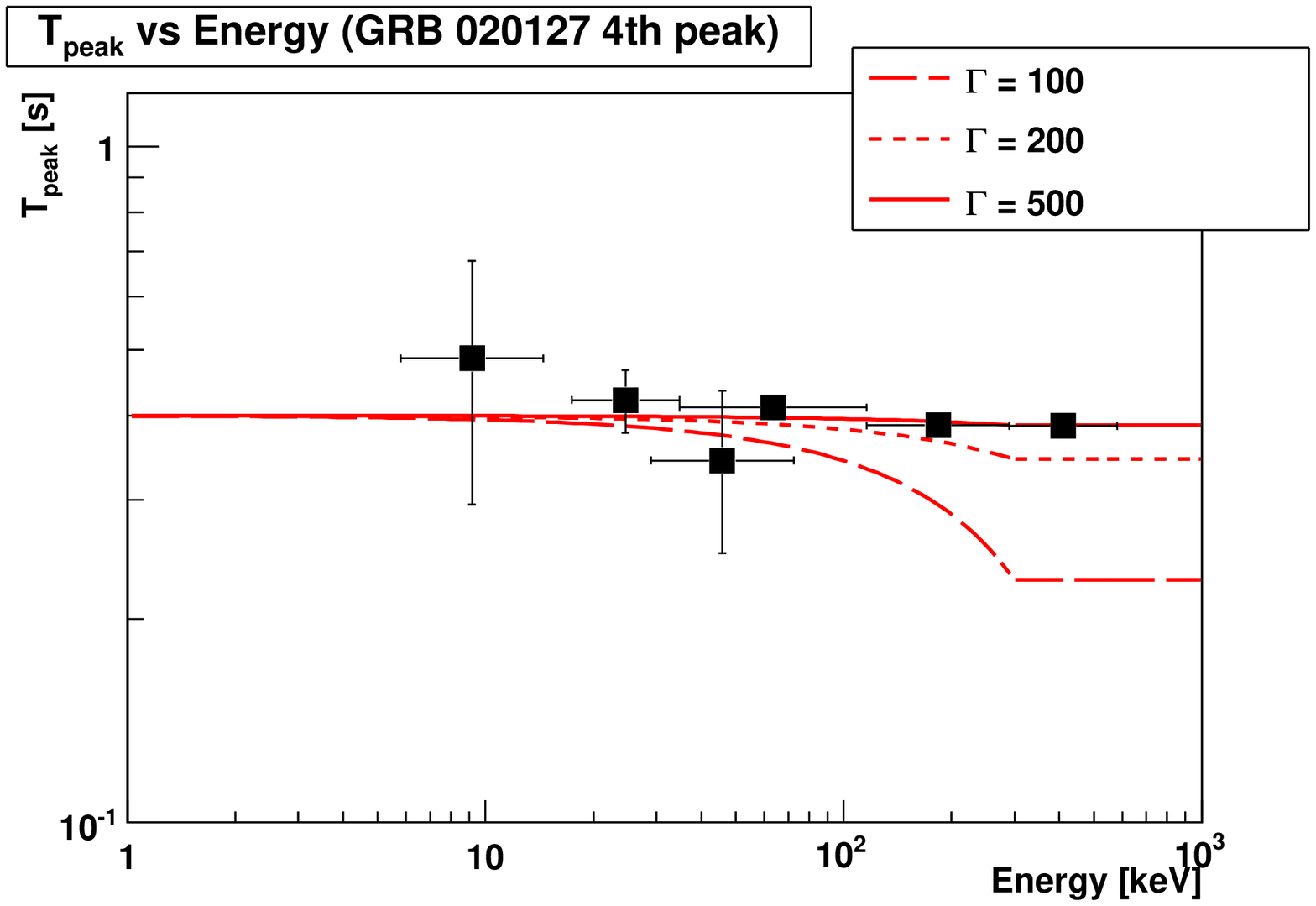}}
  \end{minipage}
 \end{center}
 \begin{center}
  \begin{minipage}[xbt]{75mm}
   \resizebox{75mm}{!}
   {\includegraphics[angle=0]{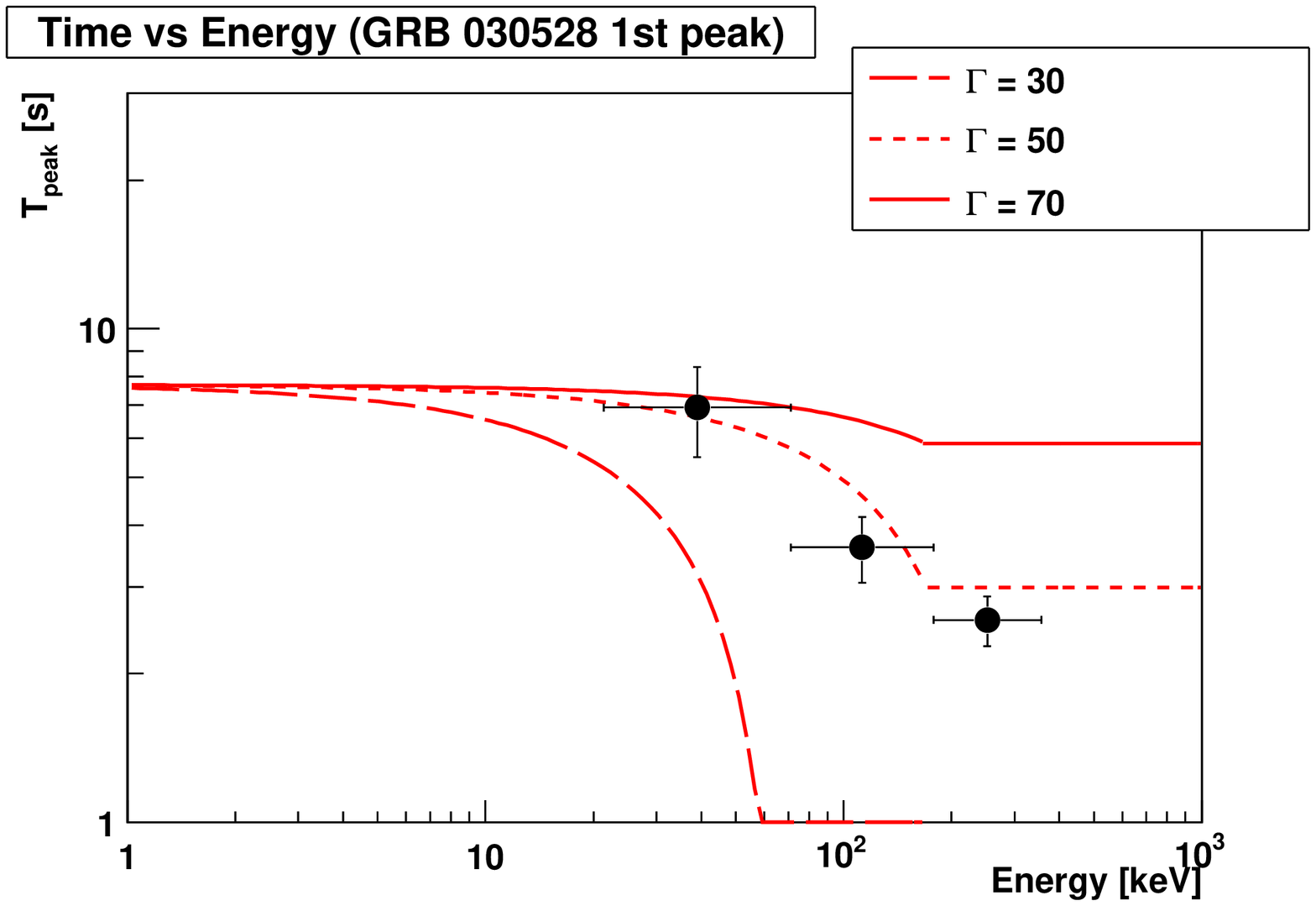}}
  \end{minipage}
  \begin{minipage}[xbt]{75mm}
   \resizebox{75mm}{!}
   {\includegraphics[angle=0]{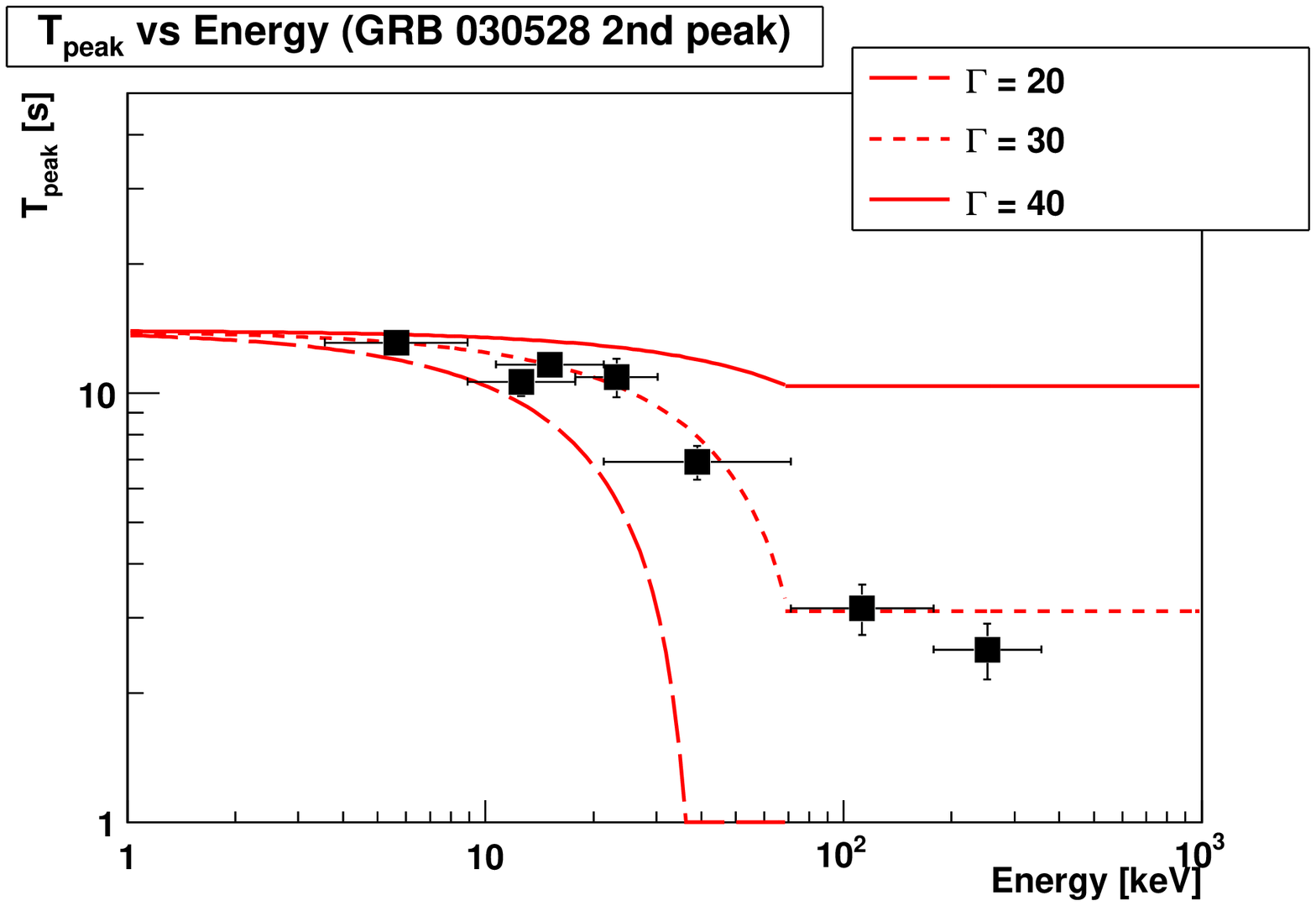}}
  \end{minipage}
 \end{center}
 \begin{center}
  \begin{minipage}[xbt]{75mm}
   \resizebox{75mm}{!}
   {\includegraphics[angle=0]{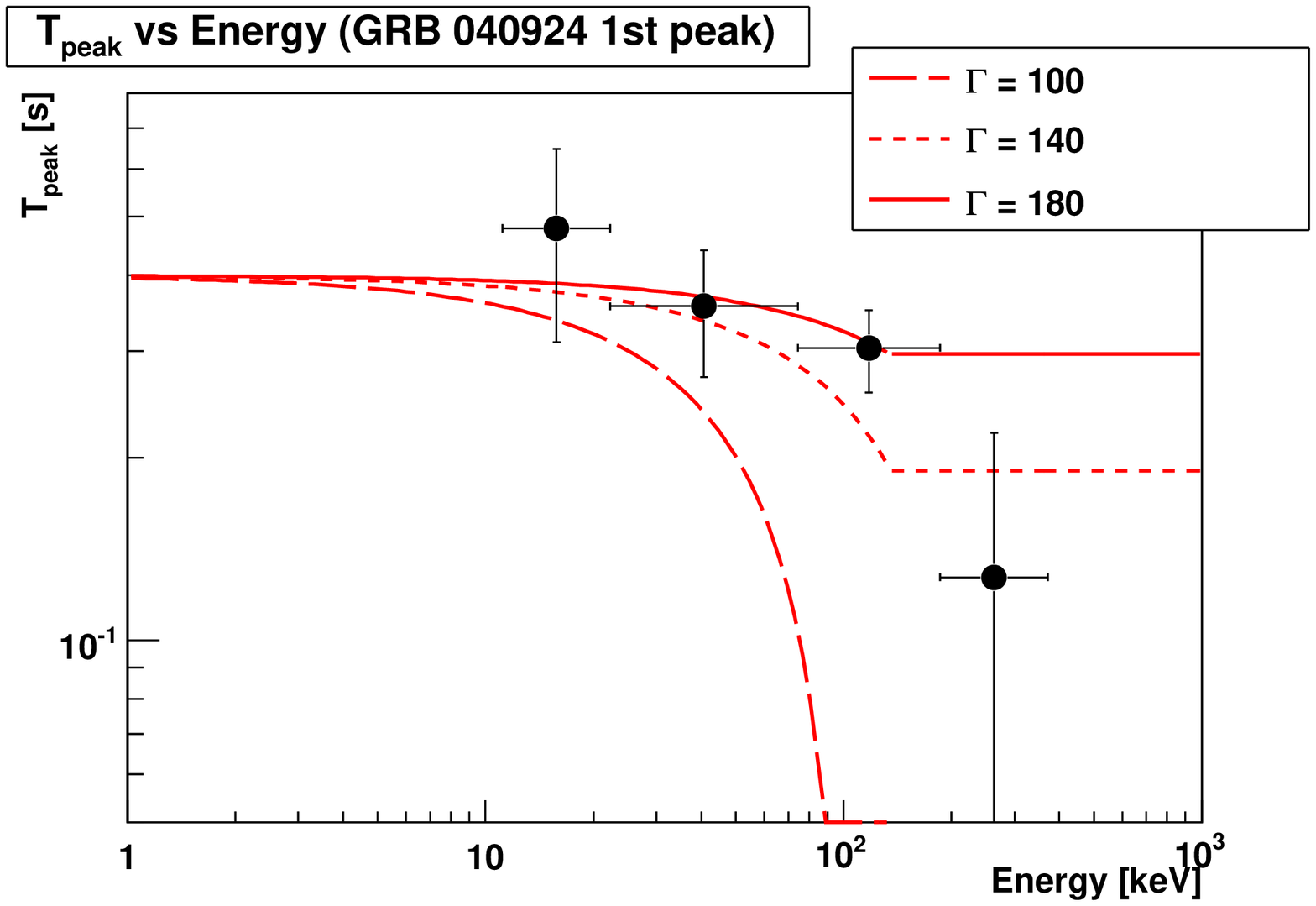}}
  \end{minipage}
  \begin{minipage}[xbt]{75mm}
   \resizebox{75mm}{!}
   {\includegraphics[angle=0]{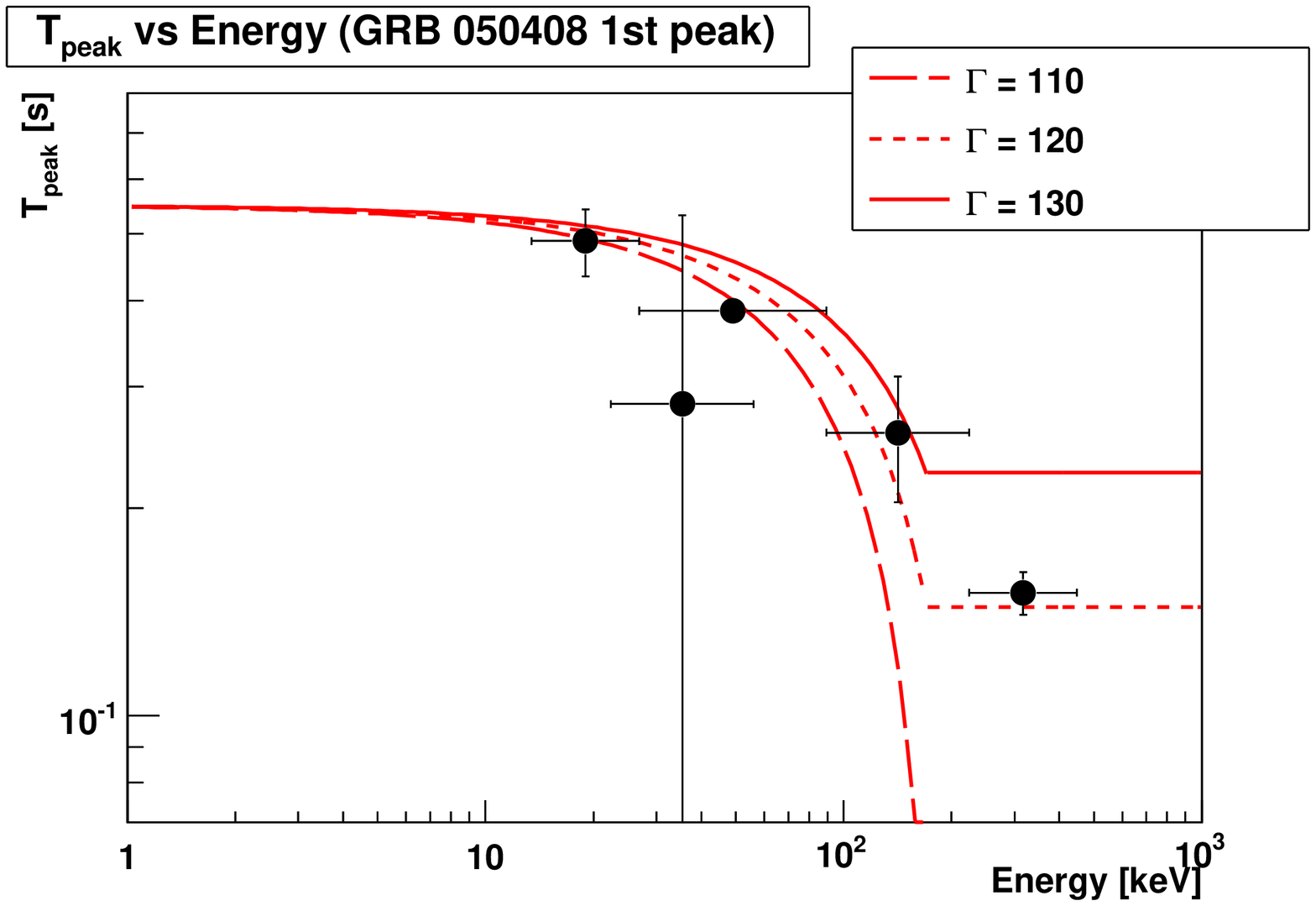}}
  \end{minipage}
 \end{center}
 \begin{center}
  \begin{minipage}[xbt]{75mm}
   \resizebox{75mm}{!}
   {\includegraphics[angle=0]{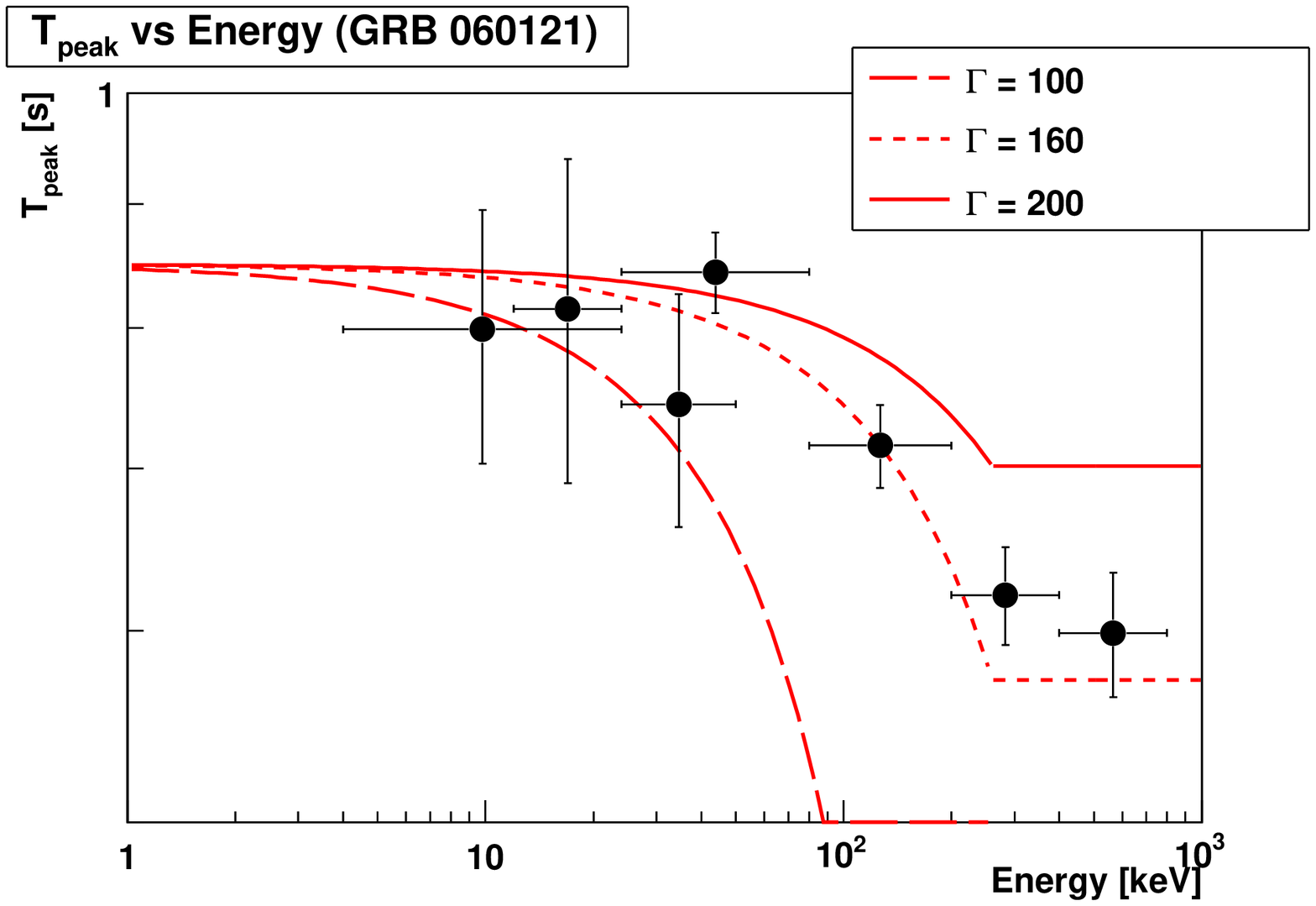}}
  \end{minipage}
  \caption{Energy vs. $T_{\rm   peak}$ plots in the burst rest frame
  with the theoretical model lines, 
showing good agreement with the curvature case. Each line represents the
  curvature-effect line at the corresponding bulk Lorentz factor $\Gamma$.}
  \label{fig:EnevsTdec_Curvature}
 \end{center}
 \end{figure*}

\begin{figure*}[htbp]
 \begin{center}
  \begin{minipage}[xbt]{75mm}
   \resizebox{75mm}{!}
   {\includegraphics[angle=0]{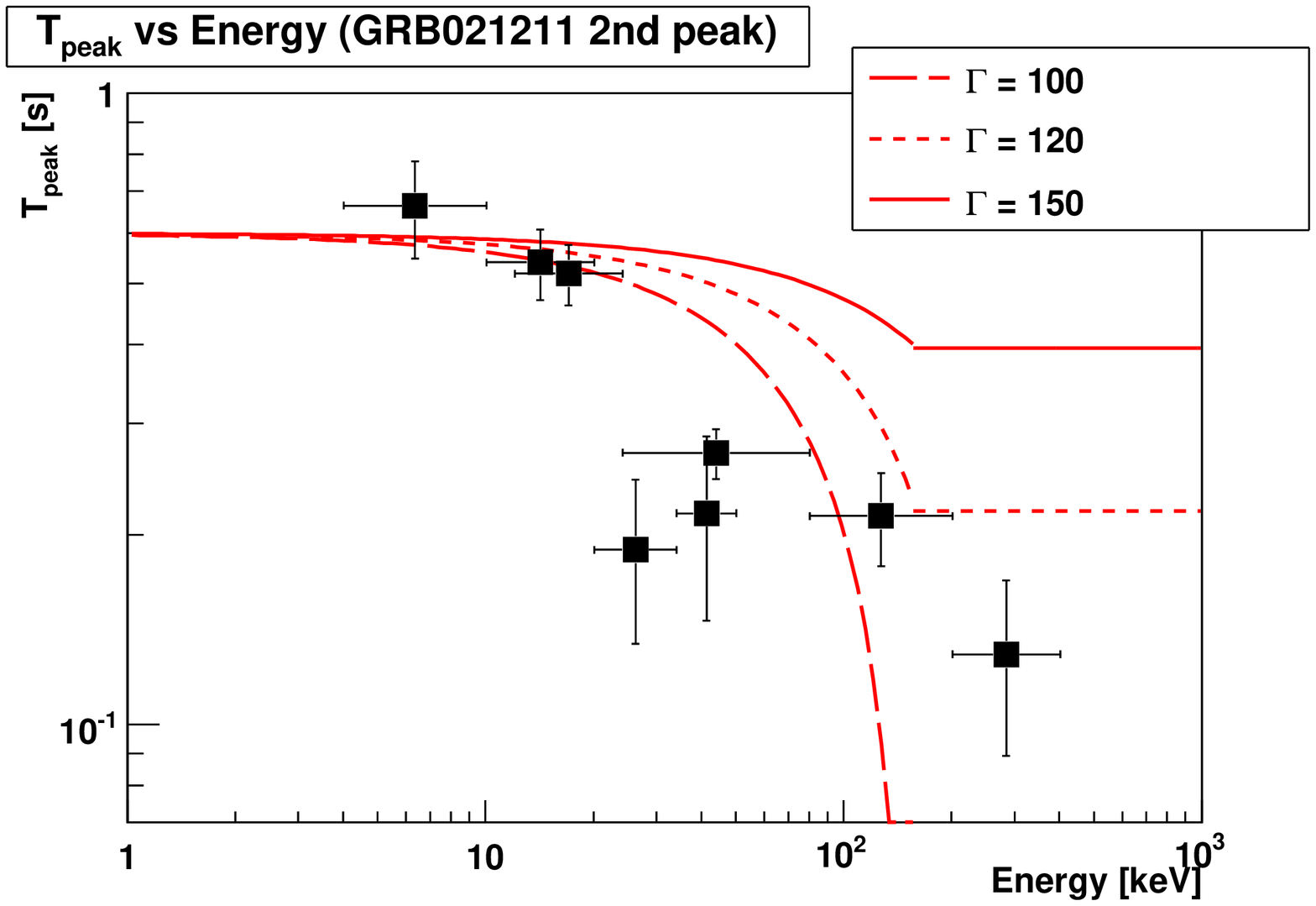}}
  \end{minipage}
  \begin{minipage}[xbt]{75mm}
   \resizebox{75mm}{!}
   {\includegraphics[angle=0]{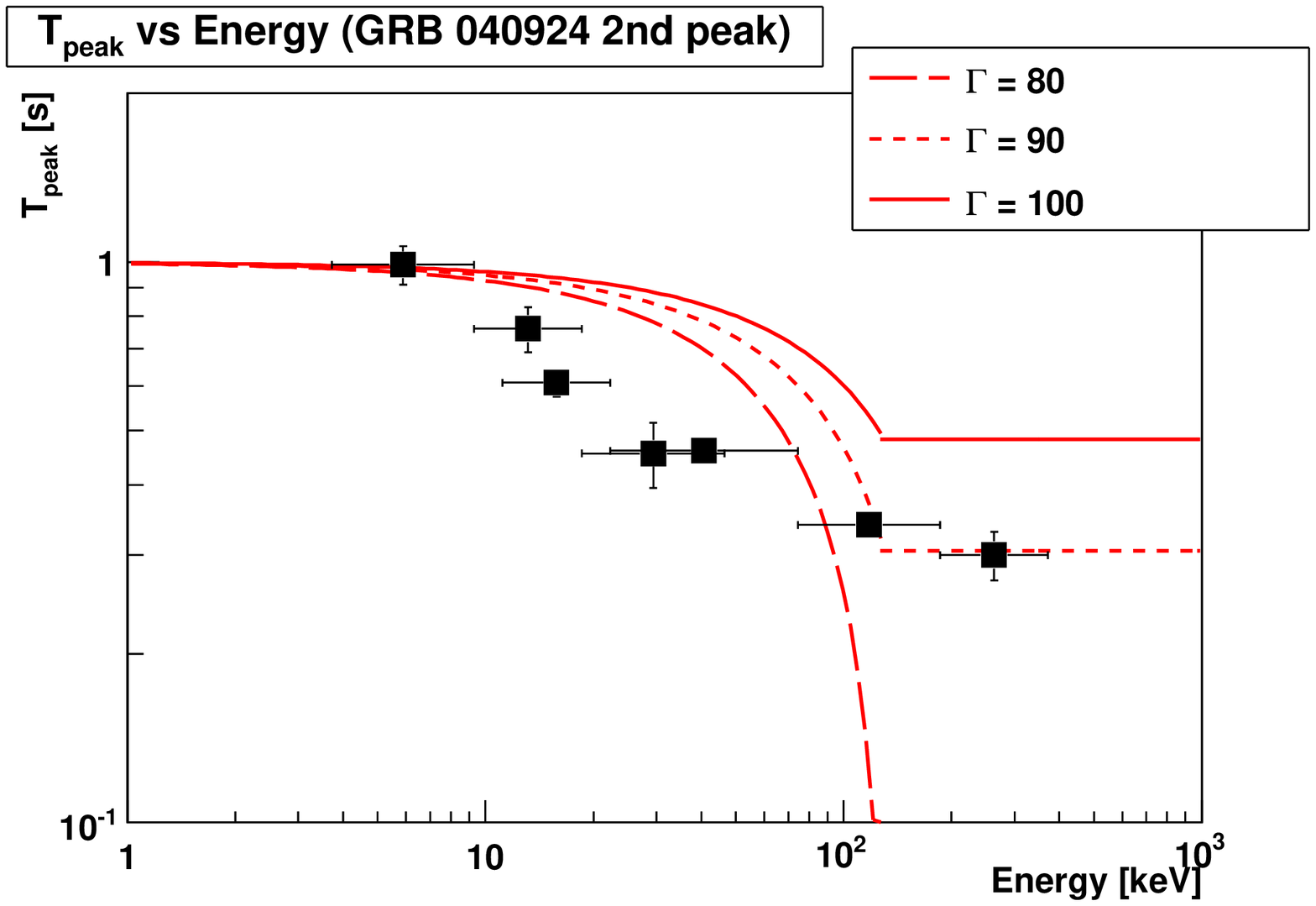}}
  \end{minipage}
 \end{center}
 \begin{center}
  \begin{minipage}[xbt]{75mm}
   \resizebox{75mm}{!}
   {\includegraphics[angle=0]{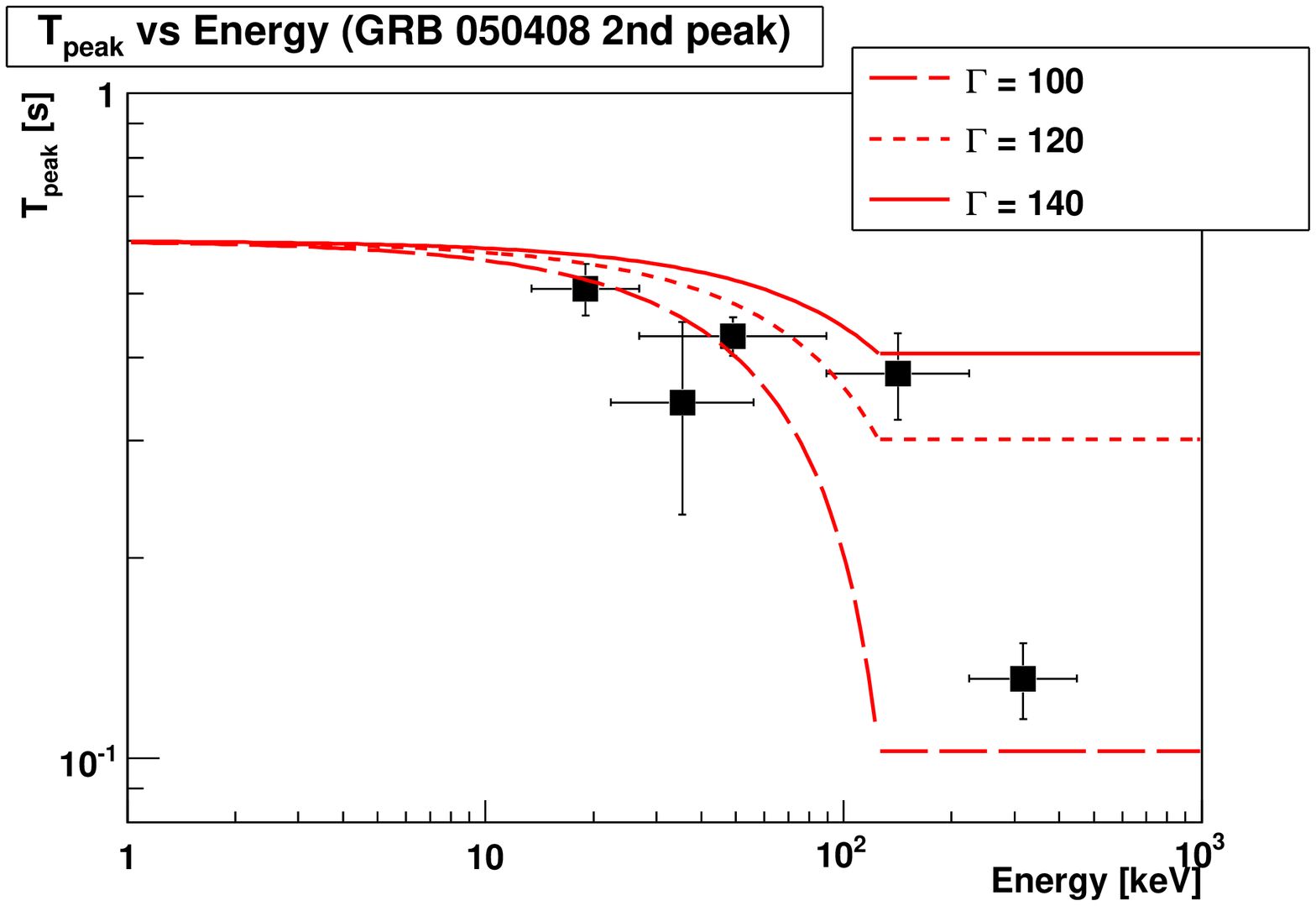}}
  \end{minipage}
  \begin{minipage}[xbt]{75mm}
   \resizebox{75mm}{!}
   {\includegraphics[angle=0]{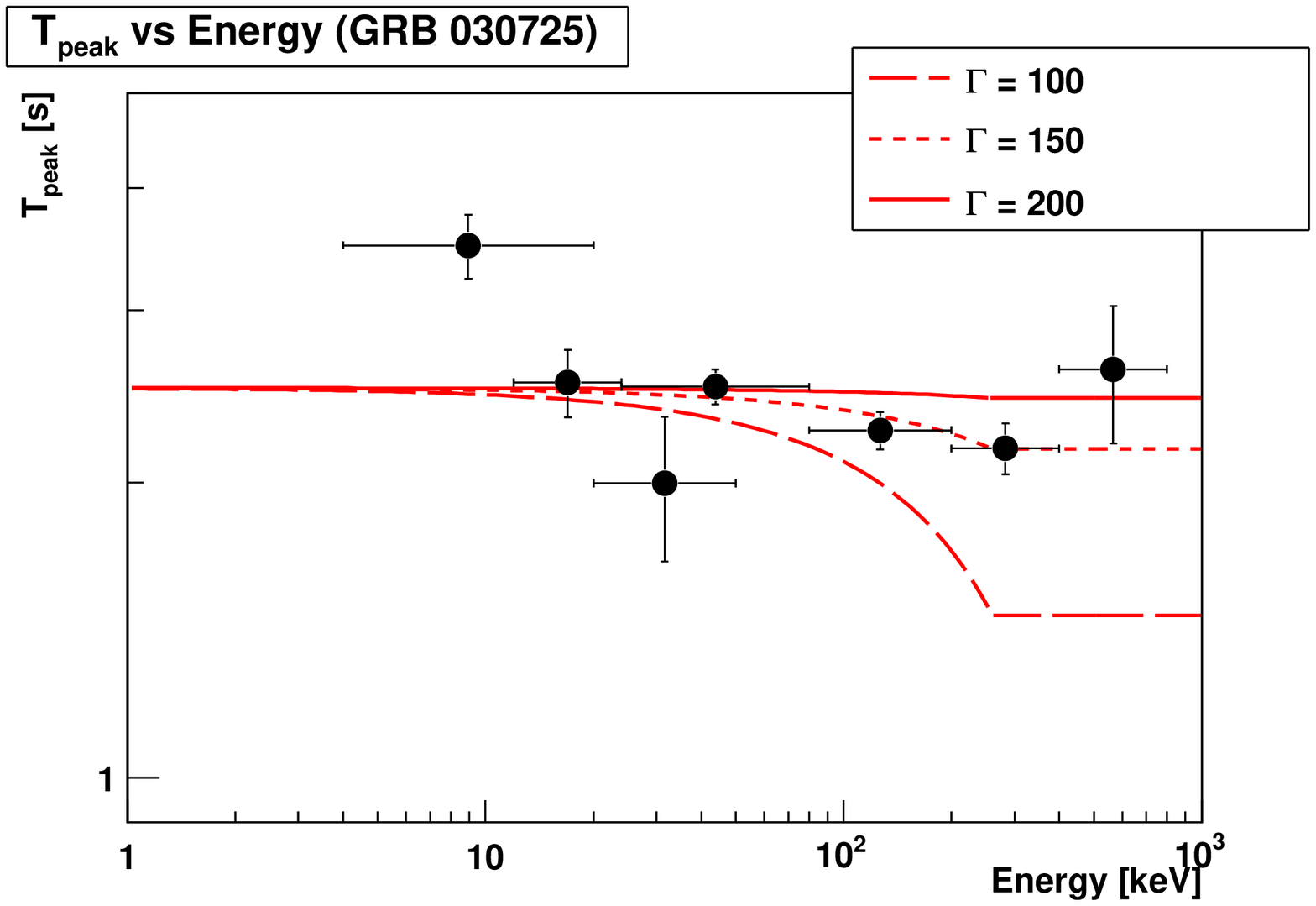}}
  \end{minipage}
  \caption{Energy vs. 
$T_{\rm peak}$ plots in the burst rest frame with the theoretical model lines, showing poor agreement with the curvature case. Each line represents the
  curvature-effect line at the corresponding bulk Lorentz factor $\Gamma$.}
  \label{fig:EnevsTdec_Hydro}
 \end{center}
 \end{figure*}

We choose the peak time at the lowest energy ($\sim$ 1 keV) arbitrarily to
match the observational value of  $T_{\rm peak}$.
\cite{2006MNRAS.367..275L} showed that the lag does not depend strongly on
the radius $r_0$. So based on their results, we write approximately
\begin{eqnarray}
\tau_{\rm lag} = \left\{
\begin{array}{ll}
a E, & \mbox{if } E \le E_s \\
a E_s, & \mbox{if } E_s < E \\
\end{array} \right. \\
\mbox{where } a = 10^{-2.4} \left( \frac{\Gamma}{100}\right)^{-2.8} \mbox{[s/keV]}. 
\end{eqnarray}
Then 
$T_{\rm peak}$ is written as
\begin{eqnarray}
 T_{\rm peak} = t_0 - \tau_{\rm lag} = \left\{
\begin{array}{ll}
t_0 - a E, & \mbox{if } E \le E_s \\
t_0 - a E_s, & \mbox{if } E_s < E \\
\end{array} \right.
\label{eq:Lu2006_2}
\end{eqnarray}
where $t_0$ is the peak time at the lowest energy. 
Using Eq. \ref{eq:Lu2006_2}, 
we try to reproduce the
spectral lag for
the examined pulses with the
curvature effect by adjusting bulk Lorentz factor $\Gamma$ 
 in Fig. \ref{fig:EnevsTdec_Curvature} and \ref{fig:EnevsTdec_Hydro}. 
Here, the energy and $T_{\rm peak}$ are translated into the burst rest frame ones with
known redshifts, and for GRB 030725 and GRB 060121 without known
redshifts assuming that their redshifts are 1.
Although our fits are based on an empirical formula
with a particular parameter set ($\alpha_B$ = -1, $\beta_B$ = -2.25),
the data points do not largely contradict the tendency
predicted by the curvature effect at a particular bulk Lorentz factor $\Gamma$, as shown in Fig. \ref{fig:EnevsTdec_Curvature}.
But for some pulses such as GRB 021211, the 2nd pulse of
GRB 040924, the 2nd pulse of GRB 050408 and GRB 030725,
the model we adopt cannot reproduce the spectral lag well
as shown in Fig. \ref{fig:EnevsTdec_Hydro}.
Even for such pulses changing  the parameters that we have fixed here
may soften the contradictions.
Alternatively,  the off-axis model or
the temporal evolution of the internal shock propagation
may play an important role in the spectral lag, or a pulse-overlap effect
may be included.

To examine whether the pulse duration and
spectral lag are explained synthetically by the curvature effect,
we summarize the results of the estimated $\Gamma$, $\alpha_{\rm w}$
 and other properties in Table \ref{table:anal_curvature_hydro}.
The empirical formula based on \cite{2006MNRAS.367..275L}
has been derived from the assumption for the intrinsic pulse duration
$\Delta t_{\rm int}$ = 10$^5$ s in the source frame.
Although there is no reason to adopt this value,
the resultant timescales estimated from the obtained $\Gamma$ are of the 
same order of magnitude as
10$^5$ s.
Even for the GRBs whose spectral lag can be explained by the curvature effect,
it is hard to confirm the consistency  between 
the prediction of the energy dependence of the pulse duration by the curvature effect
($\alpha_{ \rm w}$ = -0.2$\sim$-0.1) and the experimental values 
because of the large uncertainties in $\alpha_{\rm w}$.
However, we may say that the model based on the curvature effect does not
contradict both the spectral lag and duration at a particular bulk
Lorentz factor $\Gamma$.
Thus, the spectral lag can be a tool to help estimate
the bulk Lorentz factors.

Since in
this analysis only a finite energy range (2$-$400 keV) is available,
there are only a small number of points for the energy range, where the lag
is saturated above $E_s$.
For many GRBs, we could only plot one point above $E_s$,
 which leaves the possibility that a significant
lag takes place above $E_s$ (no saturated energy).
Even in the study by \cite{2006ApJ...653L..81L}, although the peak energy 
$E_{\rm peak}$ is $\sim$ 54 keV, they also could  plot only one point for 
$T_{\rm peak}$ above $E_{\rm peak}$ due to the poor effective area for the
higher energy ranges.
To clarify the origin of the spectral lag further, we need to detect GRB photons in the
higher energy ranges above $E_{\rm peak}$ to describe the light curve
and determine $T_{\rm peak}$ with confidence.

\begin{table*}[htb]
  \caption{Fitting parameters of the lag-luminosity -duration -$E_{\rm peak}$ relations}\label{table:param_lag-luminosity}
  \begin{center}
    \begin{tabular}{|cccccc|}
     \hline
       & & $A$ & $B$  & reduced chi-square & 
     correlation coefficient \\ \hline
     luminosity$^\dagger$ & obs & -0.79$\pm$0.04 & -1.16$\pm$0.07 & 133.2/12 & -0.79$^{+0.16}_{-0.05}$ \\ 
      & rest & -1.09$\pm$0.04 & -1.23$\pm$0.07 & 97.1/13 & -0.90$^{+0.12}_{-0.02}$ \\ \hline
     duration$^\diamond$ & obs & 1.05$\pm$0.06 & 1.16$\pm$0.09 & 43.0/12 & 0.66$^{+0.10}_{-0.14}$ \\ 
     (low energy) & rest & 1.06$\pm$0.04 & 1.15$\pm$0.06 & 72.5/13 & 0.74$^{+0.06}_{-0.15}$ \\ \hline
     duration$^{\diamond}$ & obs & 0.69$\pm$0.05 & 0.94$\pm$0.08 & 45.1/12 & 0.74$^{+0.06}_{-0.14}$ \\ 
     (high energy) & rest & 0.67$\pm$0.06 & 0.73$\pm$0.09 & 26.6/13 & 0.72$^{+0.07}_{-0.22}$ \\ \hline
     $E_{\rm peak}$$^{\ast}$ & obs & 1.88$\pm$0.02 & -0.31$\pm$0.02 & 97.8/12 & -0.66$^{+0.15}_{-0.08}$ \\ 
      & rest & 1.82$\pm$0.02 & -0.33$\pm$0.03 & 33.9/13 & -0.81$^{+0.16}_{-0.05}$ \\ \hline
      \multicolumn{6}{l}{$^\dagger$: $\log(L_{51})$ = $A + B\log(\tau_{\rm lag})$,
$^\diamond$: $\log(\omega)$ = $A + B\log(\tau_{\rm lag})$,
$^\ast$: $\log(E_{\rm peak})$ = $A + B\log(\tau_{\rm lag})$         
                }\\
      \multicolumn{6}{l}{Note that "obs" represents the observer's frame and 
        "rest" represents the rest frame.
                }\\

    \end{tabular}
  \end{center}
\end{table*}

\begin{table*}[htb]
  \caption{Summary of the obtained bulk Lorentz factor, the power-law index of 
    the duration, the observed duration and  
the expected intrinsic pulse width }\label{ttable:anal_curvature_hydro}
  \begin{center}
    \begin{tabular}{|c|c|c|c|c|}
     \hline
     GRB & $\Gamma$ & $\alpha_{\rm w}$  & $w_{\rm obs}$ [s] & 
     $\Delta t^{\rm exp}_{\rm int}$ [s] \\ \hline
     010921 & 60 & -0.11$\pm$0.18 & 21.4$\pm$2.9 & 0.77$\times$10$^{5}$  \\ 
     020127 & 500 & -0.67$\pm$0.34 & 0.70$\pm$0.01 & 1.75$\times$10$^{5}$   \\ 
     030528 (1st pulse) & 50  & -0.63$\pm$0.61 & 26.1$\pm$14.7 & 0.65$\times$10$^{5}$ \\ 
     030528 (2nd pulse) & 30  & -0.42$\pm$0.15 & 58.8$\pm$6.8  & 0.53$\times$10$^{5}$ \\ 
     040924 (1st pulse) & 140 & -0.18$\pm$0.47 & 1.8$\pm$0.5 & 0.35$\times$10$^{5}$  \\ 
     050408 (1st pulse) & 120 & -0.20$\pm$0.30 & 2.8$\pm$0.2 & 0.40$\times$10$^{5}$ \\ 
     060121             & 160 & -0.42$\pm$0.17 & 1.8$\pm$0.2 & 0.46$\times$10$^{5}$ \\ \hline
     021211             &  -  & -0.58$\pm$0.14 & 2.4$\pm$0.1 & - \\
     030725             &  -  & -0.18$\pm$0.07 & 19.7$\pm$0.5 & - \\
     040924 (2nd pulse) &  -  & -0.51$\pm$0.19 & 1.4$\pm$0.1 & -\\ 
     050408 (2nd pulse) &  -  & -0.22$\pm$0.36 & 2.0$\pm$0.2 & -\\ \hline
    \end{tabular}
   \label{table:anal_curvature_hydro}
  \end{center}
 Note that the bulk Lorentz factor $\Gamma$ is estimated by the lag analysis in
 Fig. \ref{fig:EnevsTdec_Curvature} and  $w_{\rm obs}$ is the observed
 duration in the burst rest frame and 
 $\Delta  t^{\rm exp}_{\rm int}$ is the expected intrinsic pulse width from the
obtained bulk Lorentz factor $\Gamma$.
\end{table*}

\section{Towards a  Unified Theory}
Our results suggest that there are correlations between the spectral lag
and other observational properties for each GRB pulse.
 The correlations we found
and some additional pulse correlations such as spectral hardness or
pulse asymmetry etc. reported by \cite{2009AIPC.1133..379H}
are important hints to specify or constrain models of GRB prompt emission.

On the other hand, because of the large dispersions,
the spectral lag relations are not so useful as tools
to measure cosmological distances so far, 
 compared with the Yonetoku relation.
We should note that there may still be systematic uncertainties
in the obtained lags, which may change the correlations as discussed
in \S 4.3.2.
While the obtained lag-luminosity, -$E_{\rm peak}$ and -duration
relations can be consistent with a specific model,
namely the off-axis model suggested by \cite{2001ApJ...554L.163I},
the energy dependences of the spectral lag seem to be consistent
with the simple curvature effects for some GRB pulses.
The assumptions inferred in \cite{2001ApJ...554L.163I} and
\cite{2006MNRAS.367..275L} are different so that
we have discussed the correlations and energy-dependences
in the spectral lag with two independent models.
Although such methods do not give us a consistent picture
for the spectral lag so far, the discussion in this paper may help to determine
which models are more appropriate.

For a unified theory to explain the spectral lag and other temporal
spectral characteristics, the effect of the curvature, viewing with
an offset angle to the jet, time-evolution of shock propagation,
and other effects must be taken into account synthetically and theoretical 
investigations need to be done.
To have further quantitative discussions, we need
a sample which includes many GRBs having a good S/N ratio 
detected in a wide band 
(keV$-$GeV) with observationally known redshifts.

\vspace{1pc}
 We appreciate the referee, Prof. Jon Hakkila, 
for his fruitful comments, which have improved 
our paper.
 M. A. acknowledges the financial support from the Global Center
of Excellence Program by MEXT, Japan through the Nanoscience and
Quantum Physics Project of the Tokyo Institute of Technology.
This work has been supported by Japanese Grant-in-Aid for
Young Scientists (B) 20740102.  G. P. acknowledges financial support 
a part of ASI contract I/088/06/0.


\end{document}